\documentclass[aps,prl,preprint,superscriptaddress]{revtex4-1}
\usepackage{graphicx}
\usepackage{multirow}
\newcommand{\term}[4]{\ensuremath{^{#1}#2^{#3}_{#4}}}

\begin{document}
\title{Observation of Electric-Dipole Transitions in the Laser-Cooling Candidate Th$^{-}$ }
\author{Rulin Tang}
\thanks{These authors contribute equally.}
\affiliation{Department of Physics, State Key Laboratory of Low-Dimensional Quantum Physics, Tsinghua University, Beijing 100084, China}
\author{Ran Si}
\thanks{These authors contribute equally.}
\affiliation{Division of Mathematical Physics, Department of Physics, Lund University, P.O. Box 118, 221 00 Lund, Sweden}
\affiliation{Shanghai EBIT Lab, Key Laboratory of Nuclear Physics and Ion-beam Application (MOE), Institute of Modern Physics, Department of Nuclear Science and Technology, Fudan University, Shanghai 200433, China}
\author{Zejie Fei}
\affiliation{Key Laboratory of Interfacial Physics and Technology, Shanghai Institute of Applied Physics, Chinese Academy of Sciences, Shanghai 201800, China}
\author{Xiaoxi Fu}
\affiliation{Department of Physics, State Key Laboratory of Low-Dimensional Quantum Physics, Tsinghua University, Beijing 100084, China}
\author{Yuzhu Lu}
\affiliation{Department of Physics, State Key Laboratory of Low-Dimensional Quantum Physics, Tsinghua University, Beijing 100084, China}
\author{Tomas Brage}
\affiliation{Division of Mathematical Physics, Department of Physics, Lund University, P.O. Box 118, 221 00 Lund, Sweden}
\affiliation{Shanghai EBIT Lab, Key Laboratory of Nuclear Physics and Ion-beam Application (MOE), Institute of Modern Physics, Department of Nuclear Science and Technology, Fudan University, Shanghai 200433, China}
\author{Hongtao Liu}
\email{liuhongtao@sinap.ac.cn (HTL)}
\affiliation{Key Laboratory of Interfacial Physics and Technology, Shanghai Institute of Applied Physics, Chinese Academy of Sciences, Shanghai 201800, China}
\author{Chongyang Chen}
\email{chychen@fudan.edu.cn (CYC)}
\affiliation{Shanghai EBIT Lab, Key Laboratory of Nuclear Physics and Ion-beam Application (MOE), Institute of Modern Physics, Department of Nuclear Science and Technology, Fudan University, Shanghai 200433, China}
\author{Chuangang Ning}
\email{ningcg@tsinghua.edu.cn (CGN)}
\affiliation{Department of Physics, State Key Laboratory of Low-Dimensional Quantum Physics, Tsinghua University, Beijing 100084, China}
\affiliation{Collaborative Innovation Center of Quantum Matter, Beijing 100084, China}
\date{\today}

\begin{abstract}
Despite the fact that the laser cooling method is a well-established technique to obtain ultra-cold neutral atoms and atomic cations, it has so far never been applied to atomic anions due to the lack of suitable electric-dipole transitions. Efforts of more than a decade currently have La$^{-}$ as the only promising candidate for laser cooling. Our previous work [Tang et al., Phys. Rev. Lett. 123, 203002(2019)] showed that Th$^{-}$ is also a potential candidate. Here we report on a combination of experimental and theoretical studies to determine the relevant transition frequencies, transition rates, and branching ratios in Th$^{-}$. The resonant frequency of the laser cooling transition is determined to be $\nu/c= 4118.0(10)\;\text{cm}^{-1}$. The transition rate is calculated as $A=1.17\times 10^{4}\;\text{s}^{-1}$. The branching fraction to dark states is very small, $1.47\times 10^{-10}$, thus this represents an ideal closed cycle for laser cooling. Since Th has zero nuclear spin, it is an excellent candidate to be used to sympathetically cool antiprotons in a Penning trap. 
\end{abstract}

\maketitle

The achievement of Bose-Einstein condensation, precision spectroscopy, and tests of fundamental symmetries has opened a new chapter in atomic and molecular physics. The main driving force behind this achievement is the ability to cool atoms and positive ions to $\mu$K or even lower temperatures via laser cooling techniques. Although laser cooling is a well-established technique for producing ultra-cold neutral atoms and positive ions, it has not yet been achieved for negative ions. In principle, once we produce ultracold ensembles of a specific anion system, we can use them to sympathetically cool any anions, ranging from elementary particles to molecular anions, which will promote the research of cold plasma\cite{Langin:2019cd}, ultracold chemistry\cite{Jin:2012ky}, and fundamental-physics tests\cite{Amoretti:2002hb, Ahmadi:2016id, Gerber:2018gk, Kellerbauer:2008im, Ahmadi:2017jr, Doser:2012jj}.
In contrast to neutral atoms and positive ions, which have an infinite number of bound states, negative ions have only a single bound state in most cases. The reason is that in atomic anions, the excess electron is bound mainly via polarization and correlation effects\cite{Andersen:2004jwa}. The potential experienced by this extra electron is shallow and of short range, and can therefore usually not possess bound excited states\cite{Andersen:2004jwa, Andersen:1999kx}. There are a few exceptions to this rule with atomic anions having bound excited states. It is even more rare that these bound states are of opposite parity\cite{Andersen:2004jwa, Pan:2010fr}, which can give rise to electric dipole ($E$1) transitions. As a matter of fact, there are so far only three reported $E$1 observations in atomic anions, for Os$^{-}$\cite{Bilodeau:2000td, Kellerbauer:2006ifa, Warring:2009kt, Fischer:2010gh, Pan:2010fr}, Ce$^{-}$\cite{Walter:2007iq, Walter:2011dm}, and La$^{-}$\cite{Pan:2010fr, Walter:2014fu, Jordan:2015ki, Cerchiari:2018ck}, where only La$^{-}$ is a promising candidate for laser cooling\cite{Cerchiari:2018ck}. The frequency and rate of the laser cooling transition was determined to be $\nu= 96.592\;713(91)\;\text{THz}$\cite{Jordan:2015ki, Cerchiari:2018ck} and $A=4.90(50)\times 10^{4}\;\text{s}^{-1}$\cite{Cerchiari:2018ck}, respectively. There are two major obstacles in using La$^{-}$ for cooling. Firstly, the dark states involved in the cooling transition cycle \term{3}{F}{e}{2} $\leftrightarrow$ \term{3}{D}{o}{1} give a population of metastable states. As an example, during the period of laser cooling an ensemble of La$^{-}$ ions from $100\;\text{K}$ to Doppler temperature $T_{D}= 0.17\;\mu\text{K}$, roughly 40\% of La$^{-}$ will end up in the metastable state \term{3}{F}{e}{3} with a lifetime of $132\;\text{s}$\cite{Cerchiari:2018ck}. Secondly, the nuclear spin of $^{139}$La$^{-}$ is 7/2, resulting in five hyperfine structure levels within the ground state \term{3}{F}{e}{2} of $^{139}$La$^{-}$ (with $F=11/2,9/2,7/2,5/2,3/2$), three for the excited state \term{3}{D}{o}{1} (with $F=9/2,7/2,5/2$), and seven for the metastable state \term{3}{F}{e}{3} (with $F=13/2,11/2,9/2,7/2,5/2,3/2,1/2$). Since several hyperfine levels involved in the cooling cycle are dark states, repumping laser beams are required to close the transition cycle\cite{Jordan:2015ki}.

Recently, we pointed out that Th$^{-}$ is also a potential candidate for laser cooling based on results from high-resolution photoelectron energy spectroscopy and highly accurate theoretical calculations\cite{Tang:2019ty}. The electron affinity (EA) of Th$^{-}$ was determined to be $4901.35(48)\;\text{cm}^{-1}$ or 0.607 690(60) eV. The transition for laser cooling was identified as \term{4}{F}{e}{3/2}$\leftrightarrow$ \term{2}{S}{o}{1/2} in Th$^{-}$ with a transition energy of $3904\;\text{cm}^{-1}$, and a relatively fast transition rate of $A= 1.17\times 10^{4}\;\text{s}^{-1}$. Since the Th isotope 232 has zero nuclear spin and therefore no hyperfine structure, it introduces much less potential complications for laser-cooling than La$^{-}$. This advantage is emphasized for sympathetically cooling antiprotons via laser-cooled anions. Both La$^{-}$ and the molecular anion C$_{2}^{-}$ have been proposed as two candidates\cite{Kellerbauer:2006ifa, Yzombard:2015bw, Gerber:2018gk}, but in the roadmap towards producing cold antihydrogens, the decelerated antiprotons are trapped and precooled in a Penning trap\cite{Doser:2012jj, Ahmadi:2017jr}, where a strong magnetic field of a few Tesla is used to confine antiprotons. The simpler structure of energy levels of Th$^{-}$ makes it possible to be laser cooled in this kind of traps. In a magnetic field, the ground state \term{4}{F}{e}{3/2} and the excited state \term{2}{S}{o}{1/2} are split up into four and two sublevels, respectively, due to the Zeeman effects. This is in sharp contrast to the hyperfine affected system of La$^{-}$, which will become very complex in a magnetic field. The challenge of laser-cooling molecular anions, such as C$_{2}^{-}$, is to recycle vibrational and rotational branchings of the cooling transition\cite{Yzombard:2015bw, Gerber:2018gk}. Obviously, the laser cooling of Th$^{-}$ in a Penning trap is more attractive than in a Paul trap because the precooled antiproton can be transferred and trapped efficiently using the same magnetic field. 

In this letter, we report the experimental observation of the bound-bound electrical dipole transitions in Th$^{-}$ from the ground state \term{4}{F}{e}{3/2} to excited states \term{2}{S}{o}{1/2}, \term{4}{F}{o}{5/2}, and \term{4}{D}{o}{1/2} by the resonant two-photon detachment method. We experimentally determine the resonance frequencies and obtain two-photon detachment photoelectron spectra at the resonant frequencies. Based on the previous theoretical calculations, we further extend the search for all possible bound states of Th$^{-}$. Moreover, to address the question of to which degree the cooling cycle is closed, all relevant branching ratios of transitions are deduced. 

The experiment is conducted using our cryogenic slow electron velocity-mapping imaging (cryo-SEVI) spectrometer\cite{Osterwalder:2004cx, Hock:2012fl, Wang:2008dz}, which is described in detail in earlier publications\cite{Tang:2018er}. The slow electron velocity-mapping imaging (SEVI) method has a high energy resolution for low-kinetic-energy electrons. We have used this method to determine the electron affinity (EA) of several transition elements\cite{Fu:2017ba, Chen:2016dk, Luo:2016ez}, such as Re\cite{Chen:2017ht}, Hf\cite{Tang:2018ct} and La\cite{Lu:2019df}. Th$^{-}$ ions are produced by laser sputtering on a pure thorium metal disk. Generated anions lose their kinetic energy via collisions with the buffer gas and are trapped in a radio-frequency (RF) octupole ion trap, which is mounted on a cryogenically cold head with a controlled temperature in the rang 5$-$300 K. In this experiment, the mixture of 20\% H$_{2}$ and 80\% He is used as buffer gas, which is delivered by a pulsed valve. Th$^{-}$ anions are stored in the trap for a period of 45 ms, and the temperature is kept at 300 K. Under the experimental conditions, our experimental results show that all excited Th$^{-}$ decay to the ground state. The trapped anions are then extracted via pulsed potentials on the end caps of the ion trap and analyzed by a Wiley-McLaren type time-of-flight (TOF) mass spectrometry\cite{Wiley:1955bl}. Using a mass gate, we can select Th$^{-}$ anions via a setting of $m=232$. Next, a probing laser beam with an adjustable wavelength intersects the ion beam orthogonally and photodetaches Th$^{-}$. The emitted electrons form a spherical shell and are projected onto a phosphor screen by the electric field of the velocity-map imaging system\cite{Leon:2014kj}. Each bright spot fired by a photoelectron on the phosphor screen is measured and its position is recorded with an event-counting mode via a CCD camera. Since the probing laser beam is linearly polarized parallel to the phosphor screen, the distribution of photoelectrons has cylindrical symmetry. Hence, the 3D photoelectron spherical shell can be reconstructed from the projected 2D distribution without losing information. We use the maximum-entropy reconstruction method\cite{Dick:2014jf} to reconstruct the distribution of photoelectrons. The corresponding binding energy (BE) of the detachment channel is extracted from $\text{BE} = h\nu-\alpha r^{2}$, where $h\nu$ is the photon energy, $r$ is the radius of the spherical shell, and $\alpha$ is a calibration coefficient, which can be determined by changing $h\nu$. 

To observe the $E$1 transitions in Th$^{-}$, we have recently modified the imaging system of the spectrometer, making it possible to switch from the standard SEVI mode to the scanning mode. In the scanning mode, the phosphor screen is used as a charged particle detector. A high-speed oscilloscope is connected to the phosphor screen to record both the photoelectron signals and the residual Th$^{-}$ signals after photodetachment. Due to the smaller mass of the photoelectrons, their arrival time is earlier than that of the Th$^{-}$ anions. Therefore, one channel can be used to record both signals. Since the one-color laser is used both for the resonant absorption and the photodetachment, it is possible to survey photon energies from EA/2 to EA. When the laser frequency is at the resonance, Th$^{-}$ anions can absorb one photon and reach an excited state. The excited Th$^{-}$ anions will then be detached by absorbing another photon, leading to a signal of photoelectrons. When the laser wavelength goes far from the resonance, two-photon detachment process cannot occur and there is no photoelectron. To take into account the intensity fluctuation of the Th$^{-}$ anion beam, the ratio of the intensity of the photoelectron signal to that of Th$^{-}$ anion beam is plotted versus the scanned wavelength. In both modes, the spectrometer runs at a 20-Hz repetition rate.

\begin{figure}
\includegraphics{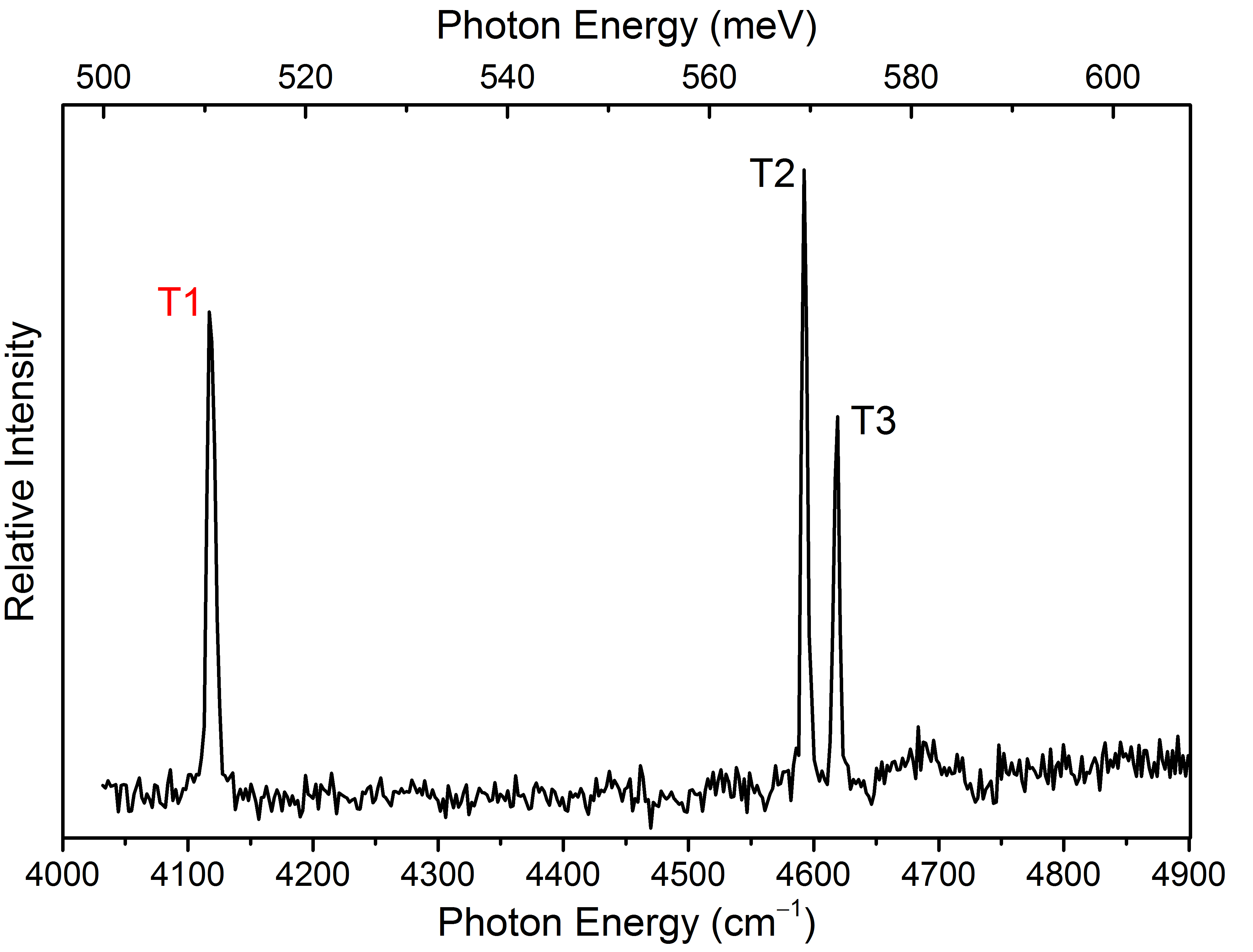}
\caption{\label{fig:1} Survey scan showing three resonances (T1, T2, and T3) in the range from 4050 to $4900\;\text{cm}^{-1}$. }
\end{figure}

To investigate the resonance we use the idle light of an OPO laser (primoScan) pumped by 355 nm, the third harmonic of the Nd:YAG (Quanta-Ray Lab 190). The idle light ranges from 700 to 2700 nm with a linewidth of about $5\;\text{cm}^{-1}$. Due to the limitation of the absorption of water vapor in the air and the signal-to-noise ratio, we performed a rough scan ranged from $4000\;\text{cm}^{-1}$ to $4900\;\text{cm}^{-1}$ with a step of $2\;\text{cm}^{-1}$ to obtain an overall spectrum. As shown in Figure \ref{fig:1}, three strong resonances were observed, labeled T1, T2, and T3. The full widths at half maximum (FWHM) of peaks are about $8\;\text{cm}^{-1}$, mainly due to the broad linewidth of the OPO laser.

\begin{figure}
\includegraphics{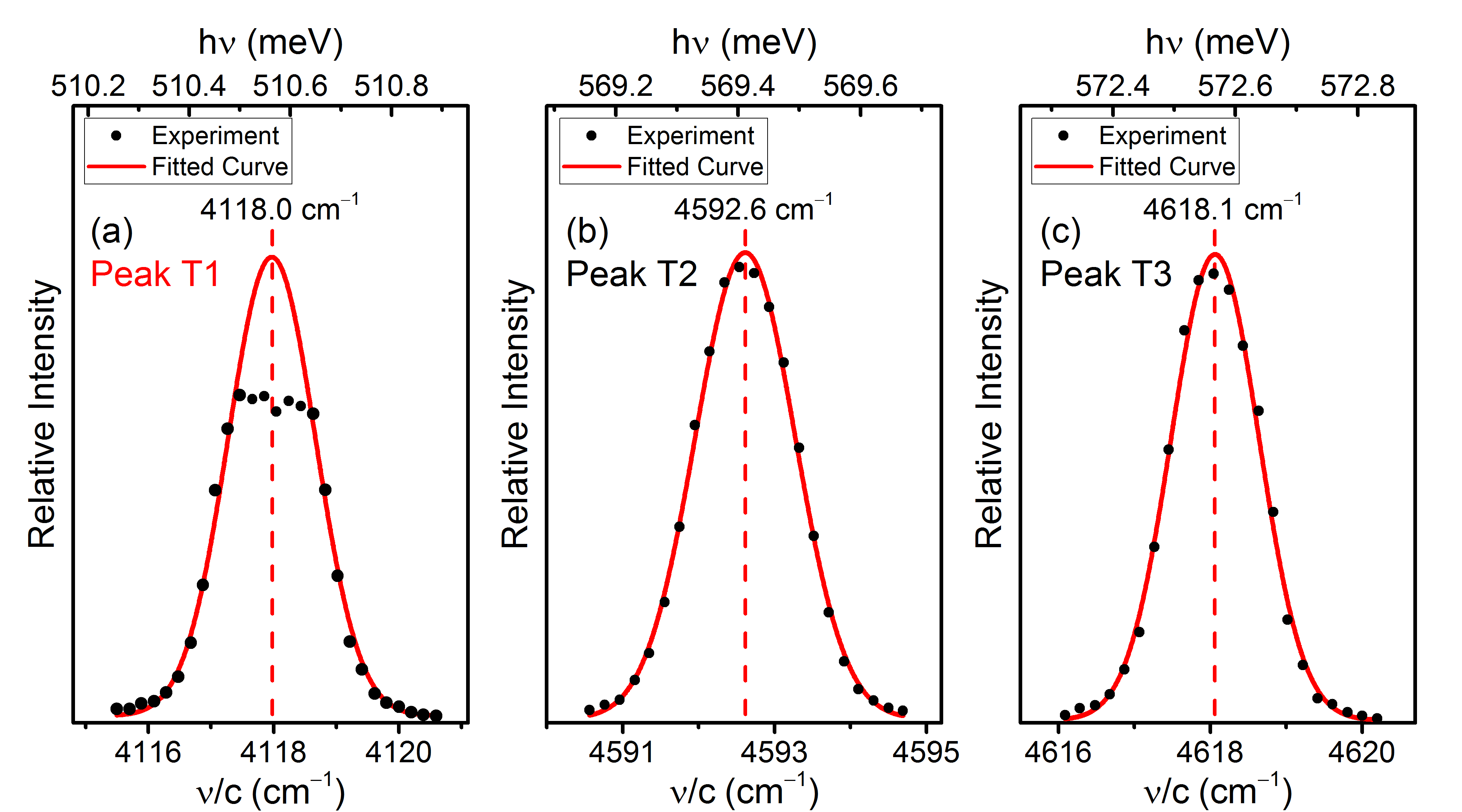}
\caption{\label{fig:2} Fine scans of the three resonances T1, T2, and T3. The solid lines indicated Gaussian fitings. Peak centers, indicated by the dashes lines, are $4118.0\;\text{cm}^{-1}$, $4592.6\;\text{cm}^{-1}$, and $4618.1\;\text{cm}^{-1}$, respectively. Peak T1 is assigned to the laser cooling transition \term{4}{F}{e}{3/2}$ \leftrightarrow$ \term{2}{S}{o}{1/2}.}
\end{figure}

To determine the resonant frequency as accurate as possible, we scan the observed resonances with a step size of only $0.2\;\text{cm}^{-1}$ using the infrared difference frequency generation (DFG) system (Sirah). The infrared laser is produced by a nonlinear DFG effect between a dye laser and a 1064 nm laser. The 1064 nm laser beam is the residual fundamental output of the pump laser. The dye laser is pumped by 532 nm, the second harmonic output of the Nd:YAG (Quanta-Ray Lab 190). Residual 1064 nm laser is mixed with the dye laser in a nonlinear LiNbO$_{3}$ crystal, producing infrared light with a frequency corresponding to the difference between the frequencies of the 1064 nm and the dye laser. The photon energies of the dye laser and the 1064 nm pumping laser are measured by a wavelength meter (HighFinesse WS6-600) with an uncertainty of $0.02\;\text{cm}^{-1}$. The linewidth of the dye laser is $0.06\;\text{cm}^{-1}$, and $\sim 1\;\text{cm}^{-1}$ for the unseeded 1064 nm laser, leading to $\sim 1\;\text{cm}^{-1}$ for the final difference-frequency light. The acquired data of each peak are then fitted to a Gaussian function, as shown in Figure \ref{fig:2}. The flat top of peak T1 is due to the saturation of resonance absorption. The resonance energies of T1, T2, and T3 are determined to be $4118.0\;\text{cm}^{-1}$, $4592.6\;\text{cm}^{-1}$ and $4618.1\;\text{cm}^{-1}$, respectively, with FWHMs of $1.6\;\text{cm}^{-1}$, $1.5\;\text{cm}^{-1}$ and $1.4\;\text{cm}^{-1}$, respectively. The widths of the three peaks mainly come from the DFG laser linewidth of about $1\;\text{cm}^{-1}$ and the power broadening. Other possible contributions, such as Doppler broadening, natural linewidth, are significantly smaller. The uncertainty of the resonance energy is estimated to be $1.0\;\text{cm}^{-1}$ (30 GHz).

To interpret the experimental results, we extended our calculations using the large-scale multi-configuration Dirac-Hartree-Fock (MCDHF) method\cite{Fischer:2016gy} as implemented in the GRASP2K package\cite{Jonsson:2013ff}. Compared with our previous calculations\cite{Tang:2019ty}, we further extend the search for all possible bound states of Th$^{-}$ anion. Three more excited states were found, in the form of $6d^{3}7s^{2}$ \term{4}{P}{e}{3/2}, $6d^{2}7s^{2}7p$ \term{4}{D}{o}{3/2}, and $6d^{2}7s^{2}7p$ \term{4}{D}{o}{1/2}, which are listed along with previous results in Table \ref{tab:1}. Decay branching fractions and transition rates of the bound states of anions are two key aspects for laser cooling. Table \ref{tab:2} lists the calculated transition rates\cite{Babushkin:1964ta, Jonsson:2007dm}, absorption line strengths\cite{Grant:1974cr}, and branching fractions for upper levels lower than and including \term{2}{S}{o}{1/2} in Th$^{-}$, including the electric-dipole ($E$1), electric-quadrupole ($E$2), magnetic-dipole ($M$1), and magnetic-quadrupole ($M$2) transitions. Details for the calculations of all transitions are summarized in the Supplemental Material\footnote{See Supplemental Material at http://link.aps.org for details for the calculations of all transitions and detailed assignments of observed peaks.}. From our results, we can deduce that the possible $E$1-allowed transitions in our scanning range are from the ground state \term{4}{F}{e}{3/2} to the excited \term{2}{S}{o}{1/2},\term{4}{F}{o}{5/2},\term{4}{D}{o}{3/2}, and \term{4}{D}{o}{1/2}. The line strength for \term{4}{F}{e}{3/2} $\rightarrow$\term{4}{D}{o}{3/2} is $1.98\times 10^{-2}$, one order of magnitude smaller than that of the other three transitions. As expected, this transition was not observed in our experiments due to the signal-to-noise limitation. 

\begin{table}
\caption{\label{tab:1}Measured and calculated excitation energies and lifetimes $\tau$ of excited states of Th$^{-}$ states.}
\begin{ruledtabular}
\begin{tabular}{cccccccc}
\multirow{2}{*}{State} & & \multicolumn{2}{c}{Measured} & & \multicolumn{3}{c}{Calculated} \\ \cline{3-4} \cline{6-8} 
            & &$\text{cm}^{-1}$     & meV     & &$\text{cm}^{-1}$   & meV  & $\tau$     \\ \hline
$6d^{3}7s^{2}$ \term{4}{F}{e}{3/2}         & &        &       & & 0    & 0   &      \\
$6d^{2}7s^{2}7p$ \term{4}{G}{o}{5/2}       & &        &       & & 401   & 50   & 51.3 ms  \\
$6d^{3}7s^{2}$ \term{4}{F}{e}{5/2}        & & 1657(6)    & 205.4(7)   & & 1377   & 171  & 0.458 s  \\
$6d^{3}7s^{2}$ \term{4}{F}{e}{7/2}        & & 2896(10)   & 359.1(12)  & & 2642   & 328  & 8.08 ms  \\
$6d^{2}7s^{2}7p$ \term{4}{F}{o}{3/2}        & &        &       & & 3033   & 376  & 15.9 ms  \\
$6d^{3}7s^{2}$ \term{4}{F}{e}{9/2}         & &        &       & & 3637   & 451  & 45.6 s   \\
$6d^{2}7s^{2}7p$ \term{2}{S}{o}{1/2}        & & 4118.0(10)  & 510.57(12)  & & 3904   & 484  & 85.5 ms  \\
$6d^{2}7s^{2}7p$ \term{4}{F}{o}{7/2}        & &        &       & & 3974   & 493  & 138 ms   \\
$6d^{2}7s^{2}7p$ \term{4}{F}{o}{5/2}        & & 4618.1(10)  & 572.57(12)  & & 3992   & 495  & 42.4 ms  \\
$6d^{3}7s^{2}$ \term{4}{P}{e}{3/2}         & &        &       & & 4284   & 531  & 274 ms   \\
$6d^{2}7s^{2}7p$ \term{4}{D}{o}{3/2}        & &        &       & & 4445   & 551  & 180 ms   \\
$6d^{2}7s^{2}7p$ \term{4}{D}{o}{1/2}        & & 4592.6(10)  & 569.41(12)  & & 4503   & 558  & 44.4 ms  \\ 
\end{tabular}
\end{ruledtabular}
\end{table}

\begin{table}
\caption{\label{tab:2}Calculated transition energies, line strengths $S$, transition rates $A$, and branching fraction of transitions except ones with an upper level higher than \term{2}{S}{o}{1/2} in Th$^{-}$. Numbers in brackets represent powers of 10. A full list can be found in the Supplemental Material\cite{Note1}. }
\begin{ruledtabular}
\begin{tabular}{ccccc}
Upper level                     & Lower level                     & $S$ (a.u.)\footnote{in atomic unit (a.u.), only electric-dipole ($E$1) transitions were considered.}  & $A$ (s$^{-1}$)    & \begin{tabular}[c]{@{}c@{}}Branching \\ fraction\end{tabular} \\ \hline
\term{4}{G}{o}{5/2} & \term{4}{F}{e}{3/2} & 8.98$[-1]$ & 1.95$[+1]$ & 1         \\
\term{4}{F}{e}{5/2} & \term{4}{F}{e}{3/2} &       & 1.00$[-1]$ & 0.046       \\
                           & \term{4}{G}{o}{5/2} & 6.63$[-3]$ & 2.08$[+0]$ & 0.954       \\
\term{4}{F}{e}{7/2} & \term{4}{F}{e}{3/2} &       & 1.44$[-5]$ & 1.16$[-7]$    \\
                           & \term{4}{G}{o}{5/2} & 4.34$[-2]$ & 1.24$[+2]$ & 0.9994       \\
                           & \term{4}{F}{e}{5/2} &       & 7.74$[-2]$ & $0.000\;625$      \\
\term{4}{F}{o}{3/2} & \term{4}{F}{e}{3/2} & 4.43$[0]$ & 6.26$[+4]$ & 0.9968       \\
                           & \term{4}{G}{o}{5/2} &       & 2.58$[-3]$ & 4.10$[-8]$    \\
                           & \term{4}{F}{e}{5/2} & 8.72$[-2]$ & 2.01$[+2]$ & $0.003\;195$      \\
                           & \term{4}{F}{e}{7/2} &       & 1.96$[-13]$ & 3.13$[-18]$   \\
\term{4}{F}{e}{9/2} & \term{4}{G}{o}{5/2} &       & 9.62$[-9]$ & 4.38$[-7]$    \\
                           & \term{4}{F}{e}{5/2} &       & 2.39$[-6]$ & $0.000\;11$      \\
                           & \term{4}{F}{e}{7/2} &       & 2.19$[-2]$ & $0.999\;89$      \\
\term{2}{S}{o}{1/2} & \term{4}{F}{e}{3/2} & 1.94$[-1]$ & 1.17$[+4]$ & $0.999\;999\;98$     \\
                           & \term{4}{G}{o}{5/2} &       & 1.72$[-6]$ & 1.47$[-10]$   \\
                           & \term{4}{F}{e}{5/2} &       & 2.10$[-9]$ & 1.80$[-13]$   \\
                           & \term{4}{F}{o}{3/2} &       & 2.32$[-4]$ & 1.98$[-8]$   
\end{tabular}
\end{ruledtabular}
\end{table}

\begin{figure}
\includegraphics{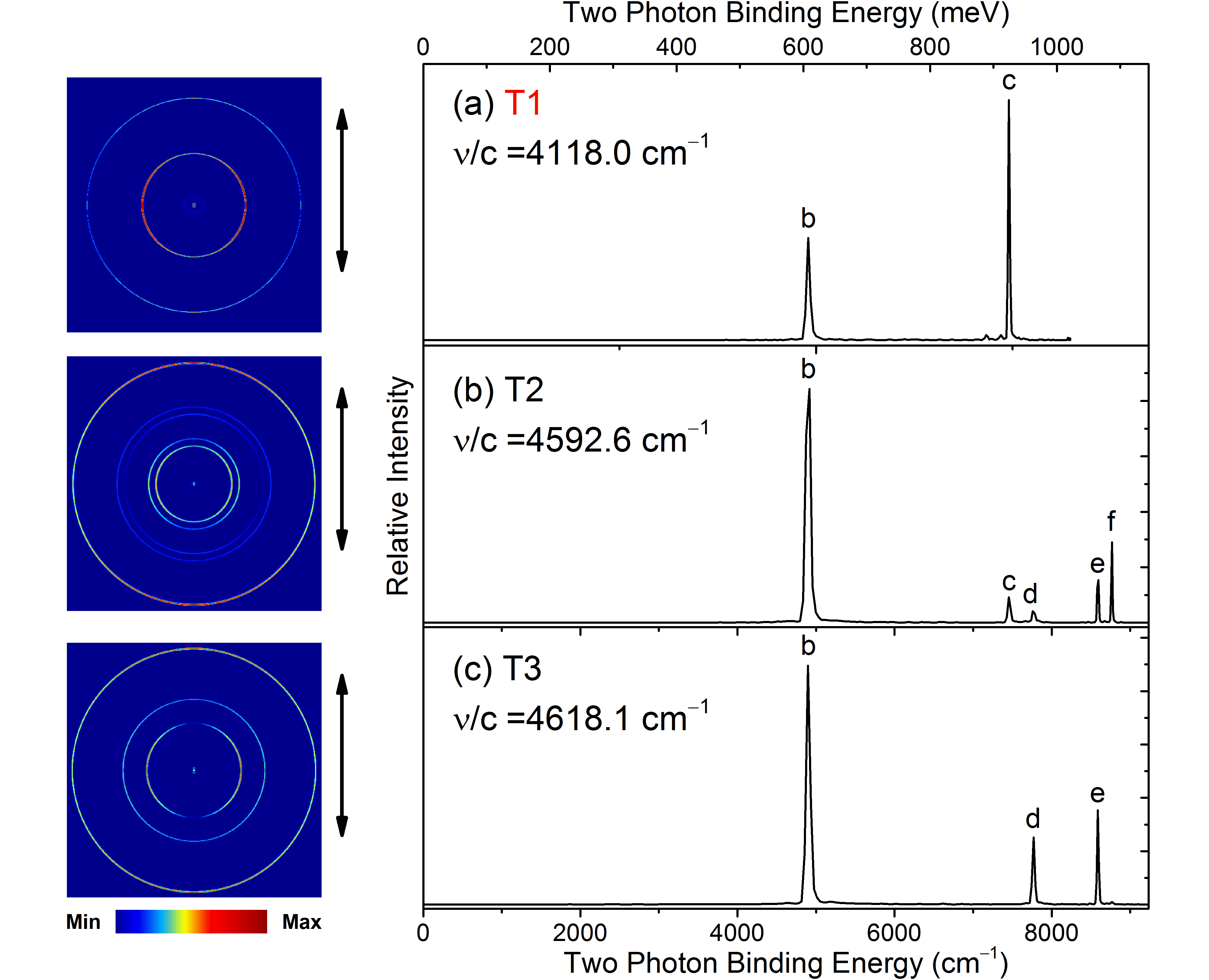}
\caption{\label{fig:3} High-Resolution two-photon photoelectron energy spectra and photoelectron images of Th$^{-}$ at three observed resonances. The double arrow indicates the laser polarization. The peaks labels are consistent with our previous work.}
\end{figure}

To further support the identification of the transitions responsible for the observed peaks, we can use additional information provided by the final states of the photodetachment. Figure \ref{fig:3} shows the resonant two-photon detachment photoelectron energy spectra of Th$^{-}$ at the three observed resonant energies, where we use the same labels as in our earlier and consistent one-step detachment experiment\cite{Tang:2019ty}. It is clear that the energy spectra acquired at the three resonances are quite different. Using the selection rules of photodetachment\cite{Engelking:1979gf}, it is clear that T1 can be assigned to the laser cooling transition \term{4}{F}{e}{3/2} $\rightarrow$ \term{2}{S}{o}{1/2} unambiguously. However, at the current stage, it is not possible to make definite identifications for T2 and T3, but a tentative assignment would be \term{4}{F}{e}{3/2} $\rightarrow$ \term{4}{F}{o}{5/2}, and \term{4}{F}{e}{3/2} $\rightarrow$ \term{4}{D}{o}{1/2} for T2 and T3, respectively, but it still needs further theoretical confirmation. See the Supplemental Material\cite{Note1} for the detailed assignment of observed peaks.

\begin{figure}
\includegraphics[width=0.5\linewidth]{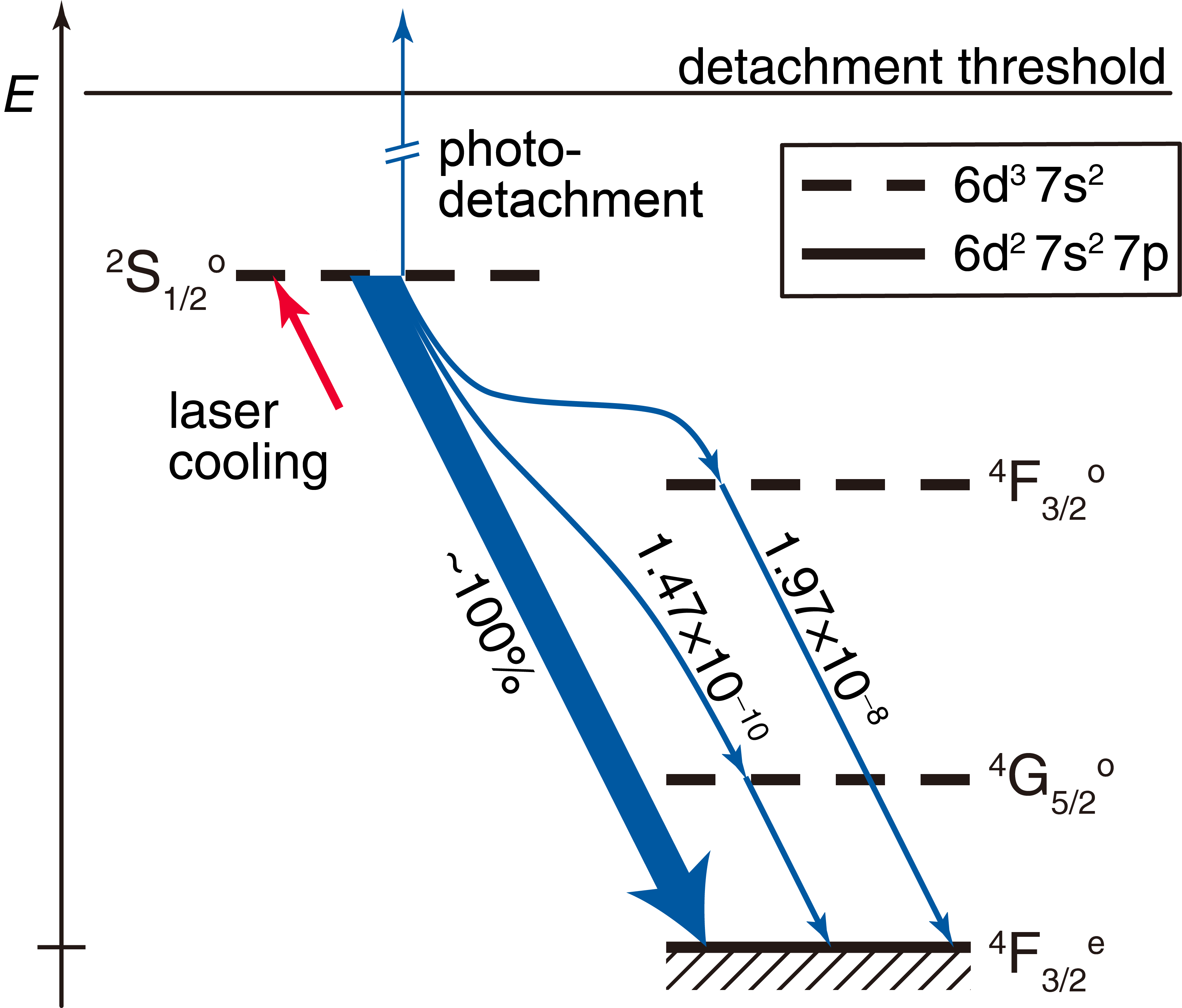}
\caption{\label{fig:4} The decay branches from the \term{2}{S}{o}{1/2} excited state. The red arrow indicates the pumping direction for laser cooling. Energies are not to scale. Thicknesses of blue arrows are indicative of branching fractions, but not to scale. }
\end{figure}

Figure \ref{fig:4} illustrates the relevant branchings of the cooling cycle. The weak transitions with branching ratio less than $10^{-10}$ can of course be neglected. We can see that almost 100\% of Th$^{-}$ \term{2}{S}{o}{1/2} decay to the ground state directly, except a ratio of $1.47\times 10^{-10}$ of Th$^{-}$ return to the ground state through \term{4}{G}{o}{5/2}, and $1.97\times 10^{-8}$ through \term{4}{F}{o}{3/2}. The lifetime of \term{4}{F}{o}{3/2} is $15.9\;\mu\text{s}$, so Th$^{-}$ anions in \term{4}{F}{o}{3/2} can quickly decay back to the ground state. \term{4}{G}{o}{5/2} has a long lifetime of 51.3 ms. Decaying to this long-life metastable state will interrupt the fast transition cycle. Therefore, \term{4}{G}{o}{5/2} is a dark state from the viewpoint of laser cooling. Fortunately, this branching ratio is very small and clearly negligible. During the period of laser cooling Th$^{-}$ from 10 K to Doppler temperature $T_{D}$, only 0.0004\% Th$^{-}$ end up in this state. No repumping laser is required. Therefore, since Th$^{-}$ does not have hyperfine structures, in principle only one laser with a wavelength $\mathcal{\lambda}=2.4284\;\text{nm}$ is required to realize the laser cooling of Th$^{-}$. The loss rate due to the photodetachment of the excited Th$^{-}$ during the cooling period is estimated to be a few percent.

In summary, from a combination of experimental and theoretical works we have shown that Th$^{-}$ is an excellent candidate for the laser cooling of anions, where the transition of laser cooling is identified as \term{4}{F}{e}{3/2} $\leftrightarrow$ \term{2}{S}{o}{1/2}. This cooling cycle is perfectly closed and the branching ended up in a dark state during the cooling period is only $4\times 10^{-6}$ relative to the cooling transition. In sharp contrast to La$^{-}$, the zero nuclear spin of Th$^{-}$ can significantly reduce the cost of the repumping laser system if the laser cooling takes place in a magnetic field. This is an important advantage for sympathetically cooling antiprotons, where a Penning trap is a more practical choice than a Paul trap to co-trap the anions and antiprotons. Before conducting the laser cooling experiment, a determination of the absolute transition rates, the photodetachment loss, and the resonant frequency of the cooling transition with an accuracy of 10 MHz is required. We will experimentally address this question in the near future. 

This work is supported by the National Natural Science Foundation of China (NSFC) (Grants No. 91736102, No. 11974199, No. 21573273, No. 11674066), the National Key R\&D Program of China (Grant No. 2018YFA0306504), Strategic Priority Research Program of the Chinese Academy of Sciences (Grant No. XDA02020000). R. S. and T.B. would like to acknowledge support of the Swedish Research Council (VR) under Contract No. 2015-04842. H.T.L would like to acknowledge support of Hundred Talents Program (CAS).

\bibliography{Th2}

\begin{thebibliography}{42}%
\makeatletter
\providecommand \@ifxundefined [1]{%
 \@ifx{#1\undefined}
}%
\providecommand \@ifnum [1]{%
 \ifnum #1\expandafter \@firstoftwo
 \else \expandafter \@secondoftwo
 \fi
}%
\providecommand \@ifx [1]{%
 \ifx #1\expandafter \@firstoftwo
 \else \expandafter \@secondoftwo
 \fi
}%
\providecommand \natexlab [1]{#1}%
\providecommand \enquote  [1]{``#1''}%
\providecommand \bibnamefont  [1]{#1}%
\providecommand \bibfnamefont [1]{#1}%
\providecommand \citenamefont [1]{#1}%
\providecommand \href@noop [0]{\@secondoftwo}%
\providecommand \href [0]{\begingroup \@sanitize@url \@href}%
\providecommand \@href[1]{\@@startlink{#1}\@@href}%
\providecommand \@@href[1]{\endgroup#1\@@endlink}%
\providecommand \@sanitize@url [0]{\catcode `\\12\catcode `\$12\catcode
  `\&12\catcode `\#12\catcode `\^12\catcode `\_12\catcode `\%12\relax}%
\providecommand \@@startlink[1]{}%
\providecommand \@@endlink[0]{}%
\providecommand \url  [0]{\begingroup\@sanitize@url \@url }%
\providecommand \@url [1]{\endgroup\@href {#1}{\urlprefix }}%
\providecommand \urlprefix  [0]{URL }%
\providecommand \Eprint [0]{\href }%
\providecommand \doibase [0]{https://doi.org/}%
\providecommand \selectlanguage [0]{\@gobble}%
\providecommand \bibinfo  [0]{\@secondoftwo}%
\providecommand \bibfield  [0]{\@secondoftwo}%
\providecommand \translation [1]{[#1]}%
\providecommand \BibitemOpen [0]{}%
\providecommand \bibitemStop [0]{}%
\providecommand \bibitemNoStop [0]{.\EOS\space}%
\providecommand \EOS [0]{\spacefactor3000\relax}%
\providecommand \BibitemShut  [1]{\csname bibitem#1\endcsname}%
\let\auto@bib@innerbib\@empty
\bibitem [{\citenamefont {Langin}\ \emph {et~al.}(2019)\citenamefont {Langin},
  \citenamefont {Gorman},\ and\ \citenamefont {Killian}}]{Langin:2019cd}%
  \BibitemOpen
  \bibfield  {author} {\bibinfo {author} {\bibfnamefont {T.~K.}\ \bibnamefont
  {Langin}}, \bibinfo {author} {\bibfnamefont {G.~M.}\ \bibnamefont {Gorman}},\
  and\ \bibinfo {author} {\bibfnamefont {T.~C.}\ \bibnamefont {Killian}},\
  }\bibfield  {title} {\bibinfo {title} {{Laser cooling of ions in a neutral
  plasma}},\ }\href@noop {} {\bibfield  {journal} {\bibinfo  {journal}
  {Science}\ }\textbf {\bibinfo {volume} {363}},\ \bibinfo {pages} {61}
  (\bibinfo {year} {2019})}\BibitemShut {NoStop}%
\bibitem [{\citenamefont {Jin}\ and\ \citenamefont {Ye}(2012)}]{Jin:2012ky}%
  \BibitemOpen
  \bibfield  {author} {\bibinfo {author} {\bibfnamefont {D.~S.}\ \bibnamefont
  {Jin}}\ and\ \bibinfo {author} {\bibfnamefont {J.}~\bibnamefont {Ye}},\
  }\bibfield  {title} {\bibinfo {title} {{Introduction to Ultracold Molecules:
  New Frontiers in Quantum and Chemical Physics}},\ }\href@noop {} {\bibfield
  {journal} {\bibinfo  {journal} {Chemical Reviews}\ }\textbf {\bibinfo
  {volume} {112}},\ \bibinfo {pages} {4801} (\bibinfo {year}
  {2012})}\BibitemShut {NoStop}%
\bibitem [{\citenamefont {Amoretti}\ \emph {et~al.}(2002)\citenamefont
  {Amoretti}, \citenamefont {Amsler}, \citenamefont {Bonomi}, \citenamefont
  {Bouchta}, \citenamefont {Bowe}, \citenamefont {Carraro}, \citenamefont
  {Cesar}, \citenamefont {Charlton}, \citenamefont {Collier}, \citenamefont
  {Doser}, \citenamefont {Filippini}, \citenamefont {Fine}, \citenamefont
  {Fontana}, \citenamefont {Fujiwara}, \citenamefont {Funakoshi}, \citenamefont
  {Genova}, \citenamefont {Hangst}, \citenamefont {Hayano}, \citenamefont
  {Holzscheiter}, \citenamefont {Jorgensen}, \citenamefont {Lagomarsino},
  \citenamefont {Landua}, \citenamefont {Lindelof}, \citenamefont {Rizzini},
  \citenamefont {Macri}, \citenamefont {Madsen}, \citenamefont {Manuzio},
  \citenamefont {Marchesotti}, \citenamefont {Montagna}, \citenamefont {Pruys},
  \citenamefont {Regenfus}, \citenamefont {Riedler}, \citenamefont {Rochet},
  \citenamefont {Rotondi}, \citenamefont {Rouleau}, \citenamefont {Testera},
  \citenamefont {Variola}, \citenamefont {Watson},\ and\ \citenamefont {van~der
  Werf}}]{Amoretti:2002hb}%
  \BibitemOpen
  \bibfield  {author} {\bibinfo {author} {\bibfnamefont {M.}~\bibnamefont
  {Amoretti}}, \bibinfo {author} {\bibfnamefont {C.}~\bibnamefont {Amsler}},
  \bibinfo {author} {\bibfnamefont {G.}~\bibnamefont {Bonomi}}, \bibinfo
  {author} {\bibfnamefont {A.}~\bibnamefont {Bouchta}}, \bibinfo {author}
  {\bibfnamefont {P.}~\bibnamefont {Bowe}}, \bibinfo {author} {\bibfnamefont
  {C.}~\bibnamefont {Carraro}}, \bibinfo {author} {\bibfnamefont {C.~L.}\
  \bibnamefont {Cesar}}, \bibinfo {author} {\bibfnamefont {M.}~\bibnamefont
  {Charlton}}, \bibinfo {author} {\bibfnamefont {M.}~\bibnamefont {Collier}},
  \bibinfo {author} {\bibfnamefont {M.}~\bibnamefont {Doser}}, \bibinfo
  {author} {\bibfnamefont {V.}~\bibnamefont {Filippini}}, \bibinfo {author}
  {\bibfnamefont {K.~S.}\ \bibnamefont {Fine}}, \bibinfo {author}
  {\bibfnamefont {A.}~\bibnamefont {Fontana}}, \bibinfo {author} {\bibfnamefont
  {M.~C.}\ \bibnamefont {Fujiwara}}, \bibinfo {author} {\bibfnamefont
  {R.}~\bibnamefont {Funakoshi}}, \bibinfo {author} {\bibfnamefont
  {P.}~\bibnamefont {Genova}}, \bibinfo {author} {\bibfnamefont {J.~S.}\
  \bibnamefont {Hangst}}, \bibinfo {author} {\bibfnamefont {R.~S.}\
  \bibnamefont {Hayano}}, \bibinfo {author} {\bibfnamefont {M.~H.}\
  \bibnamefont {Holzscheiter}}, \bibinfo {author} {\bibfnamefont {L.~V.}\
  \bibnamefont {Jorgensen}}, \bibinfo {author} {\bibfnamefont {V.}~\bibnamefont
  {Lagomarsino}}, \bibinfo {author} {\bibfnamefont {R.}~\bibnamefont {Landua}},
  \bibinfo {author} {\bibfnamefont {D.}~\bibnamefont {Lindelof}}, \bibinfo
  {author} {\bibfnamefont {E.~L.}\ \bibnamefont {Rizzini}}, \bibinfo {author}
  {\bibfnamefont {M.}~\bibnamefont {Macri}}, \bibinfo {author} {\bibfnamefont
  {N.}~\bibnamefont {Madsen}}, \bibinfo {author} {\bibfnamefont
  {G.}~\bibnamefont {Manuzio}}, \bibinfo {author} {\bibfnamefont
  {M.}~\bibnamefont {Marchesotti}}, \bibinfo {author} {\bibfnamefont
  {P.}~\bibnamefont {Montagna}}, \bibinfo {author} {\bibfnamefont
  {H.}~\bibnamefont {Pruys}}, \bibinfo {author} {\bibfnamefont
  {C.}~\bibnamefont {Regenfus}}, \bibinfo {author} {\bibfnamefont
  {P.}~\bibnamefont {Riedler}}, \bibinfo {author} {\bibfnamefont
  {J.}~\bibnamefont {Rochet}}, \bibinfo {author} {\bibfnamefont
  {A.}~\bibnamefont {Rotondi}}, \bibinfo {author} {\bibfnamefont
  {G.}~\bibnamefont {Rouleau}}, \bibinfo {author} {\bibfnamefont
  {G.}~\bibnamefont {Testera}}, \bibinfo {author} {\bibfnamefont
  {A.}~\bibnamefont {Variola}}, \bibinfo {author} {\bibfnamefont {T.~L.}\
  \bibnamefont {Watson}},\ and\ \bibinfo {author} {\bibfnamefont {D.~P.}\
  \bibnamefont {van~der Werf}},\ }\bibfield  {title} {\bibinfo {title}
  {{Production and detection of cold antihydrogen atoms}},\ }\href@noop {}
  {\bibfield  {journal} {\bibinfo  {journal} {Nature}\ }\textbf {\bibinfo
  {volume} {419}},\ \bibinfo {pages} {456} (\bibinfo {year}
  {2002})}\BibitemShut {NoStop}%
\bibitem [{\citenamefont {Ahmadi}\ \emph {et~al.}(2016)\citenamefont {Ahmadi},
  \citenamefont {Alves}, \citenamefont {Baker}, \citenamefont {Bertsche},
  \citenamefont {Butler}, \citenamefont {Capra}, \citenamefont {Carruth},
  \citenamefont {Cesar}, \citenamefont {Charlton}, \citenamefont {Cohen},
  \citenamefont {Collister}, \citenamefont {Eriksson}, \citenamefont {Evans},
  \citenamefont {Evetts}, \citenamefont {Fajans}, \citenamefont {Friesen},
  \citenamefont {Fujiwara}, \citenamefont {Gill}, \citenamefont {Gutierrez},
  \citenamefont {Hangst}, \citenamefont {Hardy}, \citenamefont {Hayden},
  \citenamefont {Isaac}, \citenamefont {Ishida}, \citenamefont {Johnson},
  \citenamefont {Jones}, \citenamefont {Jonsell}, \citenamefont {Kurchaninov},
  \citenamefont {Madsen}, \citenamefont {Mathers}, \citenamefont {Maxwell},
  \citenamefont {McKenna}, \citenamefont {Menary}, \citenamefont {Michan},
  \citenamefont {Momose}, \citenamefont {Munich}, \citenamefont {Nolan},
  \citenamefont {Olchanski}, \citenamefont {Olin}, \citenamefont {Pusa},
  \citenamefont {Rasmussen}, \citenamefont {Robicheaux}, \citenamefont
  {Sacramento}, \citenamefont {Sameed}, \citenamefont {Sarid}, \citenamefont
  {Silveira}, \citenamefont {Stracka}, \citenamefont {Stutter}, \citenamefont
  {So}, \citenamefont {Tharp}, \citenamefont {Thompson}, \citenamefont
  {Thompson}, \citenamefont {van~der Werf},\ and\ \citenamefont
  {Wurtele}}]{Ahmadi:2016id}%
  \BibitemOpen
  \bibfield  {author} {\bibinfo {author} {\bibfnamefont {M.}~\bibnamefont
  {Ahmadi}}, \bibinfo {author} {\bibfnamefont {B.~X.~R.}\ \bibnamefont
  {Alves}}, \bibinfo {author} {\bibfnamefont {C.~J.}\ \bibnamefont {Baker}},
  \bibinfo {author} {\bibfnamefont {W.}~\bibnamefont {Bertsche}}, \bibinfo
  {author} {\bibfnamefont {E.}~\bibnamefont {Butler}}, \bibinfo {author}
  {\bibfnamefont {A.}~\bibnamefont {Capra}}, \bibinfo {author} {\bibfnamefont
  {C.}~\bibnamefont {Carruth}}, \bibinfo {author} {\bibfnamefont {C.~L.}\
  \bibnamefont {Cesar}}, \bibinfo {author} {\bibfnamefont {M.}~\bibnamefont
  {Charlton}}, \bibinfo {author} {\bibfnamefont {S.}~\bibnamefont {Cohen}},
  \bibinfo {author} {\bibfnamefont {R.}~\bibnamefont {Collister}}, \bibinfo
  {author} {\bibfnamefont {S.}~\bibnamefont {Eriksson}}, \bibinfo {author}
  {\bibfnamefont {A.}~\bibnamefont {Evans}}, \bibinfo {author} {\bibfnamefont
  {N.}~\bibnamefont {Evetts}}, \bibinfo {author} {\bibfnamefont
  {J.}~\bibnamefont {Fajans}}, \bibinfo {author} {\bibfnamefont
  {T.}~\bibnamefont {Friesen}}, \bibinfo {author} {\bibfnamefont {M.~C.}\
  \bibnamefont {Fujiwara}}, \bibinfo {author} {\bibfnamefont {D.~R.}\
  \bibnamefont {Gill}}, \bibinfo {author} {\bibfnamefont {A.}~\bibnamefont
  {Gutierrez}}, \bibinfo {author} {\bibfnamefont {J.~S.}\ \bibnamefont
  {Hangst}}, \bibinfo {author} {\bibfnamefont {W.~N.}\ \bibnamefont {Hardy}},
  \bibinfo {author} {\bibfnamefont {M.~E.}\ \bibnamefont {Hayden}}, \bibinfo
  {author} {\bibfnamefont {C.~A.}\ \bibnamefont {Isaac}}, \bibinfo {author}
  {\bibfnamefont {A.}~\bibnamefont {Ishida}}, \bibinfo {author} {\bibfnamefont
  {M.~A.}\ \bibnamefont {Johnson}}, \bibinfo {author} {\bibfnamefont {S.~A.}\
  \bibnamefont {Jones}}, \bibinfo {author} {\bibfnamefont {S.}~\bibnamefont
  {Jonsell}}, \bibinfo {author} {\bibfnamefont {L.}~\bibnamefont
  {Kurchaninov}}, \bibinfo {author} {\bibfnamefont {N.}~\bibnamefont {Madsen}},
  \bibinfo {author} {\bibfnamefont {M.}~\bibnamefont {Mathers}}, \bibinfo
  {author} {\bibfnamefont {D.}~\bibnamefont {Maxwell}}, \bibinfo {author}
  {\bibfnamefont {J.~T.~K.}\ \bibnamefont {McKenna}}, \bibinfo {author}
  {\bibfnamefont {S.}~\bibnamefont {Menary}}, \bibinfo {author} {\bibfnamefont
  {J.~M.}\ \bibnamefont {Michan}}, \bibinfo {author} {\bibfnamefont
  {T.}~\bibnamefont {Momose}}, \bibinfo {author} {\bibfnamefont {J.~J.}\
  \bibnamefont {Munich}}, \bibinfo {author} {\bibfnamefont {P.}~\bibnamefont
  {Nolan}}, \bibinfo {author} {\bibfnamefont {K.}~\bibnamefont {Olchanski}},
  \bibinfo {author} {\bibfnamefont {A.}~\bibnamefont {Olin}}, \bibinfo {author}
  {\bibfnamefont {P.}~\bibnamefont {Pusa}}, \bibinfo {author} {\bibfnamefont
  {C.~{\O}.}\ \bibnamefont {Rasmussen}}, \bibinfo {author} {\bibfnamefont
  {F.}~\bibnamefont {Robicheaux}}, \bibinfo {author} {\bibfnamefont {R.~L.}\
  \bibnamefont {Sacramento}}, \bibinfo {author} {\bibfnamefont
  {M.}~\bibnamefont {Sameed}}, \bibinfo {author} {\bibfnamefont
  {E.}~\bibnamefont {Sarid}}, \bibinfo {author} {\bibfnamefont {D.~M.}\
  \bibnamefont {Silveira}}, \bibinfo {author} {\bibfnamefont {S.}~\bibnamefont
  {Stracka}}, \bibinfo {author} {\bibfnamefont {G.}~\bibnamefont {Stutter}},
  \bibinfo {author} {\bibfnamefont {C.}~\bibnamefont {So}}, \bibinfo {author}
  {\bibfnamefont {T.~D.}\ \bibnamefont {Tharp}}, \bibinfo {author}
  {\bibfnamefont {J.~E.}\ \bibnamefont {Thompson}}, \bibinfo {author}
  {\bibfnamefont {R.~I.}\ \bibnamefont {Thompson}}, \bibinfo {author}
  {\bibfnamefont {D.~P.}\ \bibnamefont {van~der Werf}},\ and\ \bibinfo {author}
  {\bibfnamefont {J.~S.}\ \bibnamefont {Wurtele}},\ }\bibfield  {title}
  {\bibinfo {title} {{Observation of the 1S{\textendash}2S transition in
  trapped antihydrogen}},\ }\href@noop {} {\bibfield  {journal} {\bibinfo
  {journal} {Nature}\ }\textbf {\bibinfo {volume} {541}},\ \bibinfo {pages}
  {506} (\bibinfo {year} {2016})}\BibitemShut {NoStop}%
\bibitem [{\citenamefont {Gerber}\ \emph {et~al.}(2018)\citenamefont {Gerber},
  \citenamefont {Fesel}, \citenamefont {Doser},\ and\ \citenamefont
  {Comparat}}]{Gerber:2018gk}%
  \BibitemOpen
  \bibfield  {author} {\bibinfo {author} {\bibfnamefont {S.}~\bibnamefont
  {Gerber}}, \bibinfo {author} {\bibfnamefont {J.}~\bibnamefont {Fesel}},
  \bibinfo {author} {\bibfnamefont {M.}~\bibnamefont {Doser}},\ and\ \bibinfo
  {author} {\bibfnamefont {D.}~\bibnamefont {Comparat}},\ }\bibfield  {title}
  {\bibinfo {title} {{Photodetachment and Doppler laser cooling of anionic
  molecules}},\ }\href@noop {} {\bibfield  {journal} {\bibinfo  {journal} {New
  Journal of Physics}\ }\textbf {\bibinfo {volume} {20}},\ \bibinfo {pages}
  {023024} (\bibinfo {year} {2018})}\BibitemShut {NoStop}%
\bibitem [{\citenamefont {Kellerbauer}\ \emph {et~al.}(2008)\citenamefont
  {Kellerbauer}, \citenamefont {Amoretti}, \citenamefont {Belov}, \citenamefont
  {Bonomi}, \citenamefont {Boscolo}, \citenamefont {Brusa}, \citenamefont
  {B{\"u}chner}, \citenamefont {Byakov}, \citenamefont {Cabaret}, \citenamefont
  {Canali}, \citenamefont {Carraro}, \citenamefont {Castelli}, \citenamefont
  {Cialdi}, \citenamefont {de~Combarieu}, \citenamefont {Comparat},
  \citenamefont {Consolati}, \citenamefont {Djourelov}, \citenamefont {Doser},
  \citenamefont {Drobychev}, \citenamefont {Dupasquier}, \citenamefont
  {Ferrari}, \citenamefont {Forget}, \citenamefont {Formaro}, \citenamefont
  {Gervasini}, \citenamefont {Giammarchi}, \citenamefont {Gninenko},
  \citenamefont {Gribakin}, \citenamefont {Hogan}, \citenamefont {Jacquey},
  \citenamefont {Lagomarsino}, \citenamefont {Manuzio}, \citenamefont
  {Mariazzi}, \citenamefont {Matveev}, \citenamefont {Meier}, \citenamefont
  {Merkt}, \citenamefont {Nedelec}, \citenamefont {Oberthaler}, \citenamefont
  {Pari}, \citenamefont {Prevedelli}, \citenamefont {Quasso}, \citenamefont
  {Rotondi}, \citenamefont {Sillou}, \citenamefont {Stepanov}, \citenamefont
  {Stroke}, \citenamefont {Testera}, \citenamefont {Tino}, \citenamefont
  {Tr{\'e}nec}, \citenamefont {Vairo}, \citenamefont {Vigu{\'e}}, \citenamefont
  {Walters}, \citenamefont {Warring}, \citenamefont {Zavatarelli},\ and\
  \citenamefont {Zvezhinskij}}]{Kellerbauer:2008im}%
  \BibitemOpen
  \bibfield  {author} {\bibinfo {author} {\bibfnamefont {A.}~\bibnamefont
  {Kellerbauer}}, \bibinfo {author} {\bibfnamefont {M.}~\bibnamefont
  {Amoretti}}, \bibinfo {author} {\bibfnamefont {A.~S.}\ \bibnamefont {Belov}},
  \bibinfo {author} {\bibfnamefont {G.}~\bibnamefont {Bonomi}}, \bibinfo
  {author} {\bibfnamefont {I.}~\bibnamefont {Boscolo}}, \bibinfo {author}
  {\bibfnamefont {R.~S.}\ \bibnamefont {Brusa}}, \bibinfo {author}
  {\bibfnamefont {M.}~\bibnamefont {B{\"u}chner}}, \bibinfo {author}
  {\bibfnamefont {V.~M.}\ \bibnamefont {Byakov}}, \bibinfo {author}
  {\bibfnamefont {L.}~\bibnamefont {Cabaret}}, \bibinfo {author} {\bibfnamefont
  {C.}~\bibnamefont {Canali}}, \bibinfo {author} {\bibfnamefont
  {C.}~\bibnamefont {Carraro}}, \bibinfo {author} {\bibfnamefont
  {F.}~\bibnamefont {Castelli}}, \bibinfo {author} {\bibfnamefont
  {S.}~\bibnamefont {Cialdi}}, \bibinfo {author} {\bibfnamefont
  {M.}~\bibnamefont {de~Combarieu}}, \bibinfo {author} {\bibfnamefont
  {D.}~\bibnamefont {Comparat}}, \bibinfo {author} {\bibfnamefont
  {G.}~\bibnamefont {Consolati}}, \bibinfo {author} {\bibfnamefont
  {N.}~\bibnamefont {Djourelov}}, \bibinfo {author} {\bibfnamefont
  {M.}~\bibnamefont {Doser}}, \bibinfo {author} {\bibfnamefont
  {G.}~\bibnamefont {Drobychev}}, \bibinfo {author} {\bibfnamefont
  {A.}~\bibnamefont {Dupasquier}}, \bibinfo {author} {\bibfnamefont
  {G.}~\bibnamefont {Ferrari}}, \bibinfo {author} {\bibfnamefont
  {P.}~\bibnamefont {Forget}}, \bibinfo {author} {\bibfnamefont
  {L.}~\bibnamefont {Formaro}}, \bibinfo {author} {\bibfnamefont
  {A.}~\bibnamefont {Gervasini}}, \bibinfo {author} {\bibfnamefont {M.~G.}\
  \bibnamefont {Giammarchi}}, \bibinfo {author} {\bibfnamefont {S.~N.}\
  \bibnamefont {Gninenko}}, \bibinfo {author} {\bibfnamefont {G.}~\bibnamefont
  {Gribakin}}, \bibinfo {author} {\bibfnamefont {S.~D.}\ \bibnamefont {Hogan}},
  \bibinfo {author} {\bibfnamefont {M.}~\bibnamefont {Jacquey}}, \bibinfo
  {author} {\bibfnamefont {V.}~\bibnamefont {Lagomarsino}}, \bibinfo {author}
  {\bibfnamefont {G.}~\bibnamefont {Manuzio}}, \bibinfo {author} {\bibfnamefont
  {S.}~\bibnamefont {Mariazzi}}, \bibinfo {author} {\bibfnamefont {V.~A.}\
  \bibnamefont {Matveev}}, \bibinfo {author} {\bibfnamefont {J.~O.}\
  \bibnamefont {Meier}}, \bibinfo {author} {\bibfnamefont {F.}~\bibnamefont
  {Merkt}}, \bibinfo {author} {\bibfnamefont {P.}~\bibnamefont {Nedelec}},
  \bibinfo {author} {\bibfnamefont {M.~K.}\ \bibnamefont {Oberthaler}},
  \bibinfo {author} {\bibfnamefont {P.}~\bibnamefont {Pari}}, \bibinfo {author}
  {\bibfnamefont {M.}~\bibnamefont {Prevedelli}}, \bibinfo {author}
  {\bibfnamefont {F.}~\bibnamefont {Quasso}}, \bibinfo {author} {\bibfnamefont
  {A.}~\bibnamefont {Rotondi}}, \bibinfo {author} {\bibfnamefont
  {D.}~\bibnamefont {Sillou}}, \bibinfo {author} {\bibfnamefont {S.~V.}\
  \bibnamefont {Stepanov}}, \bibinfo {author} {\bibfnamefont {H.~H.}\
  \bibnamefont {Stroke}}, \bibinfo {author} {\bibfnamefont {G.}~\bibnamefont
  {Testera}}, \bibinfo {author} {\bibfnamefont {G.~M.}\ \bibnamefont {Tino}},
  \bibinfo {author} {\bibfnamefont {G.}~\bibnamefont {Tr{\'e}nec}}, \bibinfo
  {author} {\bibfnamefont {A.}~\bibnamefont {Vairo}}, \bibinfo {author}
  {\bibfnamefont {J.}~\bibnamefont {Vigu{\'e}}}, \bibinfo {author}
  {\bibfnamefont {H.}~\bibnamefont {Walters}}, \bibinfo {author} {\bibfnamefont
  {U.}~\bibnamefont {Warring}}, \bibinfo {author} {\bibfnamefont
  {S.}~\bibnamefont {Zavatarelli}},\ and\ \bibinfo {author} {\bibfnamefont
  {D.~S.}\ \bibnamefont {Zvezhinskij}},\ }\bibfield  {title} {\bibinfo {title}
  {{Proposed antimatter gravity measurement with an antihydrogen beam}},\
  }\href@noop {} {\bibfield  {journal} {\bibinfo  {journal} {Nuclear
  Instruments and Methods in Physics Research Section B: Beam Interactions with
  Materials and Atoms}\ }\textbf {\bibinfo {volume} {266}},\ \bibinfo {pages}
  {351} (\bibinfo {year} {2008})}\BibitemShut {NoStop}%
\bibitem [{\citenamefont {Ahmadi}\ \emph {et~al.}(2017)\citenamefont {Ahmadi},
  \citenamefont {Alves}, \citenamefont {Baker}, \citenamefont {Bertsche},
  \citenamefont {Butler}, \citenamefont {Capra}, \citenamefont {Carruth},
  \citenamefont {Cesar}, \citenamefont {Charlton}, \citenamefont {Cohen},
  \citenamefont {Collister}, \citenamefont {Eriksson}, \citenamefont {Evans},
  \citenamefont {Evetts}, \citenamefont {Fajans}, \citenamefont {Friesen},
  \citenamefont {Fujiwara}, \citenamefont {Gill}, \citenamefont {Gutierrez},
  \citenamefont {Hangst}, \citenamefont {Hardy}, \citenamefont {Hayden},
  \citenamefont {Isaac}, \citenamefont {Ishida}, \citenamefont {Johnson},
  \citenamefont {Jones}, \citenamefont {Jonsell}, \citenamefont {Kurchaninov},
  \citenamefont {Madsen}, \citenamefont {Mathers}, \citenamefont {Maxwell},
  \citenamefont {McKenna}, \citenamefont {Menary}, \citenamefont {Michan},
  \citenamefont {Momose}, \citenamefont {Munich}, \citenamefont {Nolan},
  \citenamefont {Olchanski}, \citenamefont {Olin}, \citenamefont {Pusa},
  \citenamefont {Rasmussen}, \citenamefont {Robicheaux}, \citenamefont
  {Sacramento}, \citenamefont {Sameed}, \citenamefont {Sarid}, \citenamefont
  {Silveira}, \citenamefont {Stracka}, \citenamefont {Stutter}, \citenamefont
  {So}, \citenamefont {Tharp}, \citenamefont {Thompson}, \citenamefont
  {Thompson}, \citenamefont {van~der Werf},\ and\ \citenamefont
  {Wurtele}}]{Ahmadi:2017jr}%
  \BibitemOpen
  \bibfield  {author} {\bibinfo {author} {\bibfnamefont {M.}~\bibnamefont
  {Ahmadi}}, \bibinfo {author} {\bibfnamefont {B.~X.~R.}\ \bibnamefont
  {Alves}}, \bibinfo {author} {\bibfnamefont {C.~J.}\ \bibnamefont {Baker}},
  \bibinfo {author} {\bibfnamefont {W.}~\bibnamefont {Bertsche}}, \bibinfo
  {author} {\bibfnamefont {E.}~\bibnamefont {Butler}}, \bibinfo {author}
  {\bibfnamefont {A.}~\bibnamefont {Capra}}, \bibinfo {author} {\bibfnamefont
  {C.}~\bibnamefont {Carruth}}, \bibinfo {author} {\bibfnamefont {C.~L.}\
  \bibnamefont {Cesar}}, \bibinfo {author} {\bibfnamefont {M.}~\bibnamefont
  {Charlton}}, \bibinfo {author} {\bibfnamefont {S.}~\bibnamefont {Cohen}},
  \bibinfo {author} {\bibfnamefont {R.}~\bibnamefont {Collister}}, \bibinfo
  {author} {\bibfnamefont {S.}~\bibnamefont {Eriksson}}, \bibinfo {author}
  {\bibfnamefont {A.}~\bibnamefont {Evans}}, \bibinfo {author} {\bibfnamefont
  {N.}~\bibnamefont {Evetts}}, \bibinfo {author} {\bibfnamefont
  {J.}~\bibnamefont {Fajans}}, \bibinfo {author} {\bibfnamefont
  {T.}~\bibnamefont {Friesen}}, \bibinfo {author} {\bibfnamefont {M.~C.}\
  \bibnamefont {Fujiwara}}, \bibinfo {author} {\bibfnamefont {D.~R.}\
  \bibnamefont {Gill}}, \bibinfo {author} {\bibfnamefont {A.}~\bibnamefont
  {Gutierrez}}, \bibinfo {author} {\bibfnamefont {J.~S.}\ \bibnamefont
  {Hangst}}, \bibinfo {author} {\bibfnamefont {W.~N.}\ \bibnamefont {Hardy}},
  \bibinfo {author} {\bibfnamefont {M.~E.}\ \bibnamefont {Hayden}}, \bibinfo
  {author} {\bibfnamefont {C.~A.}\ \bibnamefont {Isaac}}, \bibinfo {author}
  {\bibfnamefont {A.}~\bibnamefont {Ishida}}, \bibinfo {author} {\bibfnamefont
  {M.~A.}\ \bibnamefont {Johnson}}, \bibinfo {author} {\bibfnamefont {S.~A.}\
  \bibnamefont {Jones}}, \bibinfo {author} {\bibfnamefont {S.}~\bibnamefont
  {Jonsell}}, \bibinfo {author} {\bibfnamefont {L.}~\bibnamefont
  {Kurchaninov}}, \bibinfo {author} {\bibfnamefont {N.}~\bibnamefont {Madsen}},
  \bibinfo {author} {\bibfnamefont {M.}~\bibnamefont {Mathers}}, \bibinfo
  {author} {\bibfnamefont {D.}~\bibnamefont {Maxwell}}, \bibinfo {author}
  {\bibfnamefont {J.~T.~K.}\ \bibnamefont {McKenna}}, \bibinfo {author}
  {\bibfnamefont {S.}~\bibnamefont {Menary}}, \bibinfo {author} {\bibfnamefont
  {J.~M.}\ \bibnamefont {Michan}}, \bibinfo {author} {\bibfnamefont
  {T.}~\bibnamefont {Momose}}, \bibinfo {author} {\bibfnamefont {J.~J.}\
  \bibnamefont {Munich}}, \bibinfo {author} {\bibfnamefont {P.}~\bibnamefont
  {Nolan}}, \bibinfo {author} {\bibfnamefont {K.}~\bibnamefont {Olchanski}},
  \bibinfo {author} {\bibfnamefont {A.}~\bibnamefont {Olin}}, \bibinfo {author}
  {\bibfnamefont {P.}~\bibnamefont {Pusa}}, \bibinfo {author} {\bibfnamefont
  {C.~{\O}.}\ \bibnamefont {Rasmussen}}, \bibinfo {author} {\bibfnamefont
  {F.}~\bibnamefont {Robicheaux}}, \bibinfo {author} {\bibfnamefont {R.~L.}\
  \bibnamefont {Sacramento}}, \bibinfo {author} {\bibfnamefont
  {M.}~\bibnamefont {Sameed}}, \bibinfo {author} {\bibfnamefont
  {E.}~\bibnamefont {Sarid}}, \bibinfo {author} {\bibfnamefont {D.~M.}\
  \bibnamefont {Silveira}}, \bibinfo {author} {\bibfnamefont {S.}~\bibnamefont
  {Stracka}}, \bibinfo {author} {\bibfnamefont {G.}~\bibnamefont {Stutter}},
  \bibinfo {author} {\bibfnamefont {C.}~\bibnamefont {So}}, \bibinfo {author}
  {\bibfnamefont {T.~D.}\ \bibnamefont {Tharp}}, \bibinfo {author}
  {\bibfnamefont {J.~E.}\ \bibnamefont {Thompson}}, \bibinfo {author}
  {\bibfnamefont {R.~I.}\ \bibnamefont {Thompson}}, \bibinfo {author}
  {\bibfnamefont {D.~P.}\ \bibnamefont {van~der Werf}},\ and\ \bibinfo {author}
  {\bibfnamefont {J.~S.}\ \bibnamefont {Wurtele}},\ }\bibfield  {title}
  {\bibinfo {title} {{Antihydrogen accumulation for fundamental symmetry
  tests}},\ }\href@noop {} {\bibfield  {journal} {\bibinfo  {journal} {Nature
  Communications}\ }\textbf {\bibinfo {volume} {8}},\ \bibinfo {pages} {681}
  (\bibinfo {year} {2017})}\BibitemShut {NoStop}%
\bibitem [{\citenamefont {Doser}\ \emph {et~al.}(2012)\citenamefont {Doser},
  \citenamefont {Amsler}, \citenamefont {Belov}, \citenamefont {Bonomi},
  \citenamefont {Br{\"a}unig}, \citenamefont {Bremer}, \citenamefont {Brusa},
  \citenamefont {Burkhart}, \citenamefont {Cabaret}, \citenamefont {Canali},
  \citenamefont {Castelli}, \citenamefont {Chlouba}, \citenamefont {Cialdi},
  \citenamefont {Comparat}, \citenamefont {Consolati}, \citenamefont {Noto},
  \citenamefont {Donzella}, \citenamefont {Dudarev}, \citenamefont {Eisel},
  \citenamefont {Ferragut}, \citenamefont {Ferrari}, \citenamefont {Fontana},
  \citenamefont {Genova}, \citenamefont {Giammarchi}, \citenamefont
  {Gligorova}, \citenamefont {Gninenko}, \citenamefont {Haider}, \citenamefont
  {Hansen}, \citenamefont {Hogan}, \citenamefont {Jorgensen}, \citenamefont
  {Kaltenbacher}, \citenamefont {Kellerbauer}, \citenamefont {Krasnicky},
  \citenamefont {Lagomarsino}, \citenamefont {Mariazzi}, \citenamefont
  {Matveev}, \citenamefont {Merkt}, \citenamefont {Moia}, \citenamefont
  {Nebbia}, \citenamefont {Nedelec}, \citenamefont {Oberthaler}, \citenamefont
  {Perini}, \citenamefont {Petracek}, \citenamefont {Prelz}, \citenamefont
  {Prevedelli}, \citenamefont {Regenfus}, \citenamefont {Riccardi},
  \citenamefont {Rohne}, \citenamefont {Rotondi}, \citenamefont {Sacerdoti},
  \citenamefont {Sandaker}, \citenamefont {Spacek}, \citenamefont {Storey},
  \citenamefont {Testera}, \citenamefont {Tokareva}, \citenamefont {Trezzi},
  \citenamefont {Vaccarone}, \citenamefont {Villa}, \citenamefont
  {Zavatarelli}, \citenamefont {Zenoni},\ and\ \citenamefont {{AEGIS
  Collaboration}}}]{Doser:2012jj}%
  \BibitemOpen
  \bibfield  {author} {\bibinfo {author} {\bibfnamefont {M.}~\bibnamefont
  {Doser}}, \bibinfo {author} {\bibfnamefont {C.}~\bibnamefont {Amsler}},
  \bibinfo {author} {\bibfnamefont {A.}~\bibnamefont {Belov}}, \bibinfo
  {author} {\bibfnamefont {G.}~\bibnamefont {Bonomi}}, \bibinfo {author}
  {\bibfnamefont {P.}~\bibnamefont {Br{\"a}unig}}, \bibinfo {author}
  {\bibfnamefont {J.}~\bibnamefont {Bremer}}, \bibinfo {author} {\bibfnamefont
  {R.}~\bibnamefont {Brusa}}, \bibinfo {author} {\bibfnamefont
  {G.}~\bibnamefont {Burkhart}}, \bibinfo {author} {\bibfnamefont
  {L.}~\bibnamefont {Cabaret}}, \bibinfo {author} {\bibfnamefont
  {C.}~\bibnamefont {Canali}}, \bibinfo {author} {\bibfnamefont
  {F.}~\bibnamefont {Castelli}}, \bibinfo {author} {\bibfnamefont
  {K.}~\bibnamefont {Chlouba}}, \bibinfo {author} {\bibfnamefont
  {S.}~\bibnamefont {Cialdi}}, \bibinfo {author} {\bibfnamefont
  {D.}~\bibnamefont {Comparat}}, \bibinfo {author} {\bibfnamefont
  {G.}~\bibnamefont {Consolati}}, \bibinfo {author} {\bibfnamefont {L.~D.}\
  \bibnamefont {Noto}}, \bibinfo {author} {\bibfnamefont {A.}~\bibnamefont
  {Donzella}}, \bibinfo {author} {\bibfnamefont {A.}~\bibnamefont {Dudarev}},
  \bibinfo {author} {\bibfnamefont {T.}~\bibnamefont {Eisel}}, \bibinfo
  {author} {\bibfnamefont {R.}~\bibnamefont {Ferragut}}, \bibinfo {author}
  {\bibfnamefont {G.}~\bibnamefont {Ferrari}}, \bibinfo {author} {\bibfnamefont
  {A.}~\bibnamefont {Fontana}}, \bibinfo {author} {\bibfnamefont
  {P.}~\bibnamefont {Genova}}, \bibinfo {author} {\bibfnamefont
  {M.}~\bibnamefont {Giammarchi}}, \bibinfo {author} {\bibfnamefont
  {A.}~\bibnamefont {Gligorova}}, \bibinfo {author} {\bibfnamefont
  {S.}~\bibnamefont {Gninenko}}, \bibinfo {author} {\bibfnamefont
  {S.}~\bibnamefont {Haider}}, \bibinfo {author} {\bibfnamefont {J.~P.}\
  \bibnamefont {Hansen}}, \bibinfo {author} {\bibfnamefont {S.}~\bibnamefont
  {Hogan}}, \bibinfo {author} {\bibfnamefont {L.}~\bibnamefont {Jorgensen}},
  \bibinfo {author} {\bibfnamefont {T.}~\bibnamefont {Kaltenbacher}}, \bibinfo
  {author} {\bibfnamefont {A.}~\bibnamefont {Kellerbauer}}, \bibinfo {author}
  {\bibfnamefont {D.}~\bibnamefont {Krasnicky}}, \bibinfo {author}
  {\bibfnamefont {V.}~\bibnamefont {Lagomarsino}}, \bibinfo {author}
  {\bibfnamefont {S.}~\bibnamefont {Mariazzi}}, \bibinfo {author}
  {\bibfnamefont {V.}~\bibnamefont {Matveev}}, \bibinfo {author} {\bibfnamefont
  {F.}~\bibnamefont {Merkt}}, \bibinfo {author} {\bibfnamefont
  {F.}~\bibnamefont {Moia}}, \bibinfo {author} {\bibfnamefont {G.}~\bibnamefont
  {Nebbia}}, \bibinfo {author} {\bibfnamefont {P.}~\bibnamefont {Nedelec}},
  \bibinfo {author} {\bibfnamefont {M.}~\bibnamefont {Oberthaler}}, \bibinfo
  {author} {\bibfnamefont {D.}~\bibnamefont {Perini}}, \bibinfo {author}
  {\bibfnamefont {V.}~\bibnamefont {Petracek}}, \bibinfo {author}
  {\bibfnamefont {F.}~\bibnamefont {Prelz}}, \bibinfo {author} {\bibfnamefont
  {M.}~\bibnamefont {Prevedelli}}, \bibinfo {author} {\bibfnamefont
  {C.}~\bibnamefont {Regenfus}}, \bibinfo {author} {\bibfnamefont
  {C.}~\bibnamefont {Riccardi}}, \bibinfo {author} {\bibfnamefont
  {O.}~\bibnamefont {Rohne}}, \bibinfo {author} {\bibfnamefont
  {A.}~\bibnamefont {Rotondi}}, \bibinfo {author} {\bibfnamefont
  {M.}~\bibnamefont {Sacerdoti}}, \bibinfo {author} {\bibfnamefont
  {H.}~\bibnamefont {Sandaker}}, \bibinfo {author} {\bibfnamefont
  {M.}~\bibnamefont {Spacek}}, \bibinfo {author} {\bibfnamefont
  {J.}~\bibnamefont {Storey}}, \bibinfo {author} {\bibfnamefont
  {G.}~\bibnamefont {Testera}}, \bibinfo {author} {\bibfnamefont
  {A.}~\bibnamefont {Tokareva}}, \bibinfo {author} {\bibfnamefont
  {D.}~\bibnamefont {Trezzi}}, \bibinfo {author} {\bibfnamefont
  {R.}~\bibnamefont {Vaccarone}}, \bibinfo {author} {\bibfnamefont
  {F.}~\bibnamefont {Villa}}, \bibinfo {author} {\bibfnamefont
  {Z.}~\bibnamefont {Zavatarelli}}, \bibinfo {author} {\bibfnamefont
  {A.}~\bibnamefont {Zenoni}},\ and\ \bibinfo {author} {\bibnamefont {{AEGIS
  Collaboration}}},\ }\bibfield  {title} {\bibinfo {title} {{Exploring the WEP
  with a pulsed cold beam of antihydrogen}},\ }\href@noop {} {\bibfield
  {journal} {\bibinfo  {journal} {Classical and Quantum Gravity}\ }\textbf
  {\bibinfo {volume} {29}},\ \bibinfo {pages} {184009} (\bibinfo {year}
  {2012})}\BibitemShut {NoStop}%
\bibitem [{\citenamefont {Andersen}(2004)}]{Andersen:2004jwa}%
  \BibitemOpen
  \bibfield  {author} {\bibinfo {author} {\bibfnamefont {T.}~\bibnamefont
  {Andersen}},\ }\bibfield  {title} {\bibinfo {title} {{Atomic negative ions:
  structure, dynamics and collisions}},\ }\href@noop {} {\bibfield  {journal}
  {\bibinfo  {journal} {Physics Reports}\ }\textbf {\bibinfo {volume} {394}},\
  \bibinfo {pages} {157} (\bibinfo {year} {2004})}\BibitemShut {NoStop}%
\bibitem [{\citenamefont {Andersen}\ \emph {et~al.}(1999)\citenamefont
  {Andersen}, \citenamefont {Haugen},\ and\ \citenamefont
  {Hotop}}]{Andersen:1999kx}%
  \BibitemOpen
  \bibfield  {author} {\bibinfo {author} {\bibfnamefont {T.}~\bibnamefont
  {Andersen}}, \bibinfo {author} {\bibfnamefont {H.~K.}\ \bibnamefont
  {Haugen}},\ and\ \bibinfo {author} {\bibfnamefont {H.}~\bibnamefont
  {Hotop}},\ }\bibfield  {title} {\bibinfo {title} {{Binding Energies in Atomic
  Negative Ions: III}},\ }\href@noop {} {\bibfield  {journal} {\bibinfo
  {journal} {Journal of Physical and Chemical Reference Data}\ }\textbf
  {\bibinfo {volume} {28}},\ \bibinfo {pages} {1511} (\bibinfo {year}
  {1999})}\BibitemShut {NoStop}%
\bibitem [{\citenamefont {Pan}\ and\ \citenamefont {Beck}(2010)}]{Pan:2010fr}%
  \BibitemOpen
  \bibfield  {author} {\bibinfo {author} {\bibfnamefont {L.}~\bibnamefont
  {Pan}}\ and\ \bibinfo {author} {\bibfnamefont {D.~R.}\ \bibnamefont {Beck}},\
  }\bibfield  {title} {\bibinfo {title} {{Candidates for laser cooling of
  atomic anions: La {-}versus Os {-}}},\ }\href@noop {} {\bibfield  {journal}
  {\bibinfo  {journal} {Physical Review A}\ }\textbf {\bibinfo {volume} {82}},\
  \bibinfo {pages} {014501} (\bibinfo {year} {2010})}\BibitemShut {NoStop}%
\bibitem [{\citenamefont {Bilodeau}\ and\ \citenamefont
  {Haugen}(2000)}]{Bilodeau:2000td}%
  \BibitemOpen
  \bibfield  {author} {\bibinfo {author} {\bibfnamefont {R.~C.}\ \bibnamefont
  {Bilodeau}}\ and\ \bibinfo {author} {\bibfnamefont {H.~K.}\ \bibnamefont
  {Haugen}},\ }\bibfield  {title} {\bibinfo {title} {{Experimental studies of
  Os-: Observation of a bound-bound electric dipole transition in an atomic
  negative ion}},\ }\href@noop {} {\bibfield  {journal} {\bibinfo  {journal}
  {Physical Review Letters}\ }\textbf {\bibinfo {volume} {85}},\ \bibinfo
  {pages} {534} (\bibinfo {year} {2000})}\BibitemShut {NoStop}%
\bibitem [{\citenamefont {Kellerbauer}\ and\ \citenamefont
  {Walz}(2006)}]{Kellerbauer:2006ifa}%
  \BibitemOpen
  \bibfield  {author} {\bibinfo {author} {\bibfnamefont {A.}~\bibnamefont
  {Kellerbauer}}\ and\ \bibinfo {author} {\bibfnamefont {J.}~\bibnamefont
  {Walz}},\ }\bibfield  {title} {\bibinfo {title} {{A novel cooling scheme for
  antiprotons}},\ }\href@noop {} {\bibfield  {journal} {\bibinfo  {journal}
  {New Journal of Physics}\ }\textbf {\bibinfo {volume} {8}},\ \bibinfo {pages}
  {45} (\bibinfo {year} {2006})}\BibitemShut {NoStop}%
\bibitem [{\citenamefont {Warring}\ \emph {et~al.}(2009)\citenamefont
  {Warring}, \citenamefont {Amoretti}, \citenamefont {Canali}, \citenamefont
  {Fischer}, \citenamefont {Heyne}, \citenamefont {Meier}, \citenamefont
  {Morhard},\ and\ \citenamefont {Kellerbauer}}]{Warring:2009kt}%
  \BibitemOpen
  \bibfield  {author} {\bibinfo {author} {\bibfnamefont {U.}~\bibnamefont
  {Warring}}, \bibinfo {author} {\bibfnamefont {M.}~\bibnamefont {Amoretti}},
  \bibinfo {author} {\bibfnamefont {C.}~\bibnamefont {Canali}}, \bibinfo
  {author} {\bibfnamefont {A.}~\bibnamefont {Fischer}}, \bibinfo {author}
  {\bibfnamefont {R.}~\bibnamefont {Heyne}}, \bibinfo {author} {\bibfnamefont
  {J.~O.}\ \bibnamefont {Meier}}, \bibinfo {author} {\bibfnamefont
  {C.}~\bibnamefont {Morhard}},\ and\ \bibinfo {author} {\bibfnamefont
  {A.}~\bibnamefont {Kellerbauer}},\ }\bibfield  {title} {\bibinfo {title}
  {{High-Resolution Laser Spectroscopy on the Negative Osmium Ion}},\
  }\href@noop {} {\bibfield  {journal} {\bibinfo  {journal} {Physical Review
  Letters}\ }\textbf {\bibinfo {volume} {102}},\ \bibinfo {pages} {043001}
  (\bibinfo {year} {2009})}\BibitemShut {NoStop}%
\bibitem [{\citenamefont {Fischer}\ \emph {et~al.}(2010)\citenamefont
  {Fischer}, \citenamefont {Canali}, \citenamefont {Warring}, \citenamefont
  {Kellerbauer},\ and\ \citenamefont {Fritzsche}}]{Fischer:2010gh}%
  \BibitemOpen
  \bibfield  {author} {\bibinfo {author} {\bibfnamefont {A.}~\bibnamefont
  {Fischer}}, \bibinfo {author} {\bibfnamefont {C.}~\bibnamefont {Canali}},
  \bibinfo {author} {\bibfnamefont {U.}~\bibnamefont {Warring}}, \bibinfo
  {author} {\bibfnamefont {A.}~\bibnamefont {Kellerbauer}},\ and\ \bibinfo
  {author} {\bibfnamefont {S.}~\bibnamefont {Fritzsche}},\ }\bibfield  {title}
  {\bibinfo {title} {{First Optical Hyperfine Structure Measurement in an
  Atomic Anion}},\ }\href@noop {} {\bibfield  {journal} {\bibinfo  {journal}
  {Physical Review Letters}\ }\textbf {\bibinfo {volume} {104}},\ \bibinfo
  {pages} {073004} (\bibinfo {year} {2010})}\BibitemShut {NoStop}%
\bibitem [{\citenamefont {Walter}\ \emph {et~al.}(2007)\citenamefont {Walter},
  \citenamefont {Gibson}, \citenamefont {Janczak}, \citenamefont {Starr},
  \citenamefont {Snedden}, \citenamefont {Field~III},\ and\ \citenamefont
  {Andersson}}]{Walter:2007iq}%
  \BibitemOpen
  \bibfield  {author} {\bibinfo {author} {\bibfnamefont {C.~W.}\ \bibnamefont
  {Walter}}, \bibinfo {author} {\bibfnamefont {N.~D.}\ \bibnamefont {Gibson}},
  \bibinfo {author} {\bibfnamefont {C.~M.}\ \bibnamefont {Janczak}}, \bibinfo
  {author} {\bibfnamefont {K.~A.}\ \bibnamefont {Starr}}, \bibinfo {author}
  {\bibfnamefont {A.~P.}\ \bibnamefont {Snedden}}, \bibinfo {author}
  {\bibfnamefont {R.~L.}\ \bibnamefont {Field~III}},\ and\ \bibinfo {author}
  {\bibfnamefont {P.}~\bibnamefont {Andersson}},\ }\bibfield  {title} {\bibinfo
  {title} {{Infrared photodetachment of Ce{-}: Threshold spectroscopy and
  resonance structure}},\ }\href@noop {} {\bibfield  {journal} {\bibinfo
  {journal} {Physical Review A}\ }\textbf {\bibinfo {volume} {76}},\ \bibinfo
  {pages} {052702} (\bibinfo {year} {2007})}\BibitemShut {NoStop}%
\bibitem [{\citenamefont {Walter}\ \emph {et~al.}(2011)\citenamefont {Walter},
  \citenamefont {Gibson}, \citenamefont {Li}, \citenamefont {Matyas},
  \citenamefont {Alton}, \citenamefont {Lou}, \citenamefont {Field},
  \citenamefont {Hanstorp}, \citenamefont {Pan},\ and\ \citenamefont
  {Beck}}]{Walter:2011dm}%
  \BibitemOpen
  \bibfield  {author} {\bibinfo {author} {\bibfnamefont {C.~W.}\ \bibnamefont
  {Walter}}, \bibinfo {author} {\bibfnamefont {N.~D.}\ \bibnamefont {Gibson}},
  \bibinfo {author} {\bibfnamefont {Y.~G.}\ \bibnamefont {Li}}, \bibinfo
  {author} {\bibfnamefont {D.~J.}\ \bibnamefont {Matyas}}, \bibinfo {author}
  {\bibfnamefont {R.~M.}\ \bibnamefont {Alton}}, \bibinfo {author}
  {\bibfnamefont {S.~E.}\ \bibnamefont {Lou}}, \bibinfo {author} {\bibfnamefont
  {R.~L.}\ \bibnamefont {Field}}, \bibinfo {author} {\bibfnamefont
  {D.}~\bibnamefont {Hanstorp}}, \bibinfo {author} {\bibfnamefont
  {L.}~\bibnamefont {Pan}},\ and\ \bibinfo {author} {\bibfnamefont {D.~R.}\
  \bibnamefont {Beck}},\ }\bibfield  {title} {\bibinfo {title} {{Experimental
  and theoretical study of bound and quasibound states of Ce{-}}},\ }\href@noop
  {} {\bibfield  {journal} {\bibinfo  {journal} {Physical Review A}\ }\textbf
  {\bibinfo {volume} {84}},\ \bibinfo {pages} {032514} (\bibinfo {year}
  {2011})}\BibitemShut {NoStop}%
\bibitem [{\citenamefont {Walter}\ \emph {et~al.}(2014)\citenamefont {Walter},
  \citenamefont {Gibson}, \citenamefont {Matyas}, \citenamefont {Crocker},
  \citenamefont {Dungan}, \citenamefont {Matola},\ and\ \citenamefont
  {Rohlen}}]{Walter:2014fu}%
  \BibitemOpen
  \bibfield  {author} {\bibinfo {author} {\bibfnamefont {C.~W.}\ \bibnamefont
  {Walter}}, \bibinfo {author} {\bibfnamefont {N.~D.}\ \bibnamefont {Gibson}},
  \bibinfo {author} {\bibfnamefont {D.~J.}\ \bibnamefont {Matyas}}, \bibinfo
  {author} {\bibfnamefont {C.}~\bibnamefont {Crocker}}, \bibinfo {author}
  {\bibfnamefont {K.~A.}\ \bibnamefont {Dungan}}, \bibinfo {author}
  {\bibfnamefont {B.~R.}\ \bibnamefont {Matola}},\ and\ \bibinfo {author}
  {\bibfnamefont {J.}~\bibnamefont {Rohlen}},\ }\bibfield  {title} {\bibinfo
  {title} {{Candidate for Laser Cooling of a Negative Ion: Observations of
  Bound-Bound Transitions in La-}},\ }\href@noop {} {\bibfield  {journal}
  {\bibinfo  {journal} {Physical Review Letters}\ }\textbf {\bibinfo {volume}
  {113}},\ \bibinfo {pages} {063001} (\bibinfo {year} {2014})}\BibitemShut
  {NoStop}%
\bibitem [{\citenamefont {Jordan}\ \emph {et~al.}(2015)\citenamefont {Jordan},
  \citenamefont {Cerchiari}, \citenamefont {Fritzsche},\ and\ \citenamefont
  {Kellerbauer}}]{Jordan:2015ki}%
  \BibitemOpen
  \bibfield  {author} {\bibinfo {author} {\bibfnamefont {E.}~\bibnamefont
  {Jordan}}, \bibinfo {author} {\bibfnamefont {G.}~\bibnamefont {Cerchiari}},
  \bibinfo {author} {\bibfnamefont {S.}~\bibnamefont {Fritzsche}},\ and\
  \bibinfo {author} {\bibfnamefont {A.}~\bibnamefont {Kellerbauer}},\
  }\bibfield  {title} {\bibinfo {title} {{High-Resolution Spectroscopy on the
  Laser-Cooling Candidate La{-}}},\ }\href@noop {} {\bibfield  {journal}
  {\bibinfo  {journal} {Physical Review Letters}\ }\textbf {\bibinfo {volume}
  {115}},\ \bibinfo {pages} {113001} (\bibinfo {year} {2015})}\BibitemShut
  {NoStop}%
\bibitem [{\citenamefont {Cerchiari}\ \emph {et~al.}(2018)\citenamefont
  {Cerchiari}, \citenamefont {Kellerbauer}, \citenamefont {Safronova},
  \citenamefont {Safronova},\ and\ \citenamefont
  {Yzombard}}]{Cerchiari:2018ck}%
  \BibitemOpen
  \bibfield  {author} {\bibinfo {author} {\bibfnamefont {G.}~\bibnamefont
  {Cerchiari}}, \bibinfo {author} {\bibfnamefont {A.}~\bibnamefont
  {Kellerbauer}}, \bibinfo {author} {\bibfnamefont {M.~S.}\ \bibnamefont
  {Safronova}}, \bibinfo {author} {\bibfnamefont {U.~I.}\ \bibnamefont
  {Safronova}},\ and\ \bibinfo {author} {\bibfnamefont {P.}~\bibnamefont
  {Yzombard}},\ }\bibfield  {title} {\bibinfo {title} {{Ultracold Anions for
  High-Precision Antihydrogen Experiments}},\ }\href@noop {} {\bibfield
  {journal} {\bibinfo  {journal} {Physical Review Letters}\ }\textbf {\bibinfo
  {volume} {120}},\ \bibinfo {pages} {133205} (\bibinfo {year}
  {2018})}\BibitemShut {NoStop}%
\bibitem [{\citenamefont {Tang}\ \emph {et~al.}(2019)\citenamefont {Tang},
  \citenamefont {Si}, \citenamefont {Fei}, \citenamefont {Fu}, \citenamefont
  {Lu}, \citenamefont {Brage}, \citenamefont {Liu}, \citenamefont {Chen},\ and\
  \citenamefont {Ning}}]{Tang:2019ty}%
  \BibitemOpen
  \bibfield  {author} {\bibinfo {author} {\bibfnamefont {R.~L.}\ \bibnamefont
  {Tang}}, \bibinfo {author} {\bibfnamefont {R.}~\bibnamefont {Si}}, \bibinfo
  {author} {\bibfnamefont {Z.}~\bibnamefont {Fei}}, \bibinfo {author}
  {\bibfnamefont {X.~X.}\ \bibnamefont {Fu}}, \bibinfo {author} {\bibfnamefont
  {Y.}~\bibnamefont {Lu}}, \bibinfo {author} {\bibfnamefont {T.}~\bibnamefont
  {Brage}}, \bibinfo {author} {\bibfnamefont {H.}~\bibnamefont {Liu}}, \bibinfo
  {author} {\bibfnamefont {C.}~\bibnamefont {Chen}},\ and\ \bibinfo {author}
  {\bibfnamefont {C.~G.}\ \bibnamefont {Ning}},\ }\bibfield  {title} {\bibinfo
  {title} {{Candidate for Laser Cooling of a Negative Ion: High-Resolution
  Photoelectron Imaging of Th{-}}},\ }\href@noop {} {\bibfield  {journal}
  {\bibinfo  {journal} {Physical Review Letters}\ }\textbf {\bibinfo {volume}
  {123}},\ \bibinfo {pages} {749} (\bibinfo {year} {2019})}\BibitemShut
  {NoStop}%
\bibitem [{\citenamefont {Yzombard}\ \emph {et~al.}(2015)\citenamefont
  {Yzombard}, \citenamefont {Hamamda}, \citenamefont {Gerber}, \citenamefont
  {Doser},\ and\ \citenamefont {Comparat}}]{Yzombard:2015bw}%
  \BibitemOpen
  \bibfield  {author} {\bibinfo {author} {\bibfnamefont {P.}~\bibnamefont
  {Yzombard}}, \bibinfo {author} {\bibfnamefont {M.}~\bibnamefont {Hamamda}},
  \bibinfo {author} {\bibfnamefont {S.}~\bibnamefont {Gerber}}, \bibinfo
  {author} {\bibfnamefont {M.}~\bibnamefont {Doser}},\ and\ \bibinfo {author}
  {\bibfnamefont {D.}~\bibnamefont {Comparat}},\ }\bibfield  {title} {\bibinfo
  {title} {{Laser Cooling of Molecular Anions}},\ }\href@noop {} {\bibfield
  {journal} {\bibinfo  {journal} {Physical Review Letters}\ }\textbf {\bibinfo
  {volume} {114}},\ \bibinfo {pages} {213001} (\bibinfo {year}
  {2015})}\BibitemShut {NoStop}%
\bibitem [{\citenamefont {Osterwalder}\ \emph {et~al.}(2004)\citenamefont
  {Osterwalder}, \citenamefont {Nee}, \citenamefont {Zhou},\ and\ \citenamefont
  {Neumark}}]{Osterwalder:2004cx}%
  \BibitemOpen
  \bibfield  {author} {\bibinfo {author} {\bibfnamefont {A.}~\bibnamefont
  {Osterwalder}}, \bibinfo {author} {\bibfnamefont {M.~J.}\ \bibnamefont
  {Nee}}, \bibinfo {author} {\bibfnamefont {J.}~\bibnamefont {Zhou}},\ and\
  \bibinfo {author} {\bibfnamefont {D.~M.}\ \bibnamefont {Neumark}},\
  }\bibfield  {title} {\bibinfo {title} {{High resolution photodetachment
  spectroscopy of negative ions via slow photoelectron imaging}},\ }\href@noop
  {} {\bibfield  {journal} {\bibinfo  {journal} {The Journal of Chemical
  Physics}\ }\textbf {\bibinfo {volume} {121}},\ \bibinfo {pages} {6317}
  (\bibinfo {year} {2004})}\BibitemShut {NoStop}%
\bibitem [{\citenamefont {Hock}\ \emph {et~al.}(2012)\citenamefont {Hock},
  \citenamefont {Kim}, \citenamefont {Weichman}, \citenamefont {Yacovitch},\
  and\ \citenamefont {Neumark}}]{Hock:2012fl}%
  \BibitemOpen
  \bibfield  {author} {\bibinfo {author} {\bibfnamefont {C.}~\bibnamefont
  {Hock}}, \bibinfo {author} {\bibfnamefont {J.~B.}\ \bibnamefont {Kim}},
  \bibinfo {author} {\bibfnamefont {M.~L.}\ \bibnamefont {Weichman}}, \bibinfo
  {author} {\bibfnamefont {T.~I.}\ \bibnamefont {Yacovitch}},\ and\ \bibinfo
  {author} {\bibfnamefont {D.~M.}\ \bibnamefont {Neumark}},\ }\bibfield
  {title} {\bibinfo {title} {{Slow photoelectron velocity-map imaging
  spectroscopy of cold negative ions}},\ }\href@noop {} {\bibfield  {journal}
  {\bibinfo  {journal} {The Journal of Chemical Physics}\ }\textbf {\bibinfo
  {volume} {137}},\ \bibinfo {pages} {244201} (\bibinfo {year}
  {2012})}\BibitemShut {NoStop}%
\bibitem [{\citenamefont {Wang}\ and\ \citenamefont
  {Wang}(2008)}]{Wang:2008dz}%
  \BibitemOpen
  \bibfield  {author} {\bibinfo {author} {\bibfnamefont {X.~B.}\ \bibnamefont
  {Wang}}\ and\ \bibinfo {author} {\bibfnamefont {L.~S.}\ \bibnamefont
  {Wang}},\ }\bibfield  {title} {\bibinfo {title} {{Development of a
  low-temperature photoelectron spectroscopy instrument using an electrospray
  ion source and a cryogenically controlled ion trap}},\ }\href@noop {}
  {\bibfield  {journal} {\bibinfo  {journal} {Review of Scientific
  Instruments}\ }\textbf {\bibinfo {volume} {79}},\ \bibinfo {pages} {073108}
  (\bibinfo {year} {2008})}\BibitemShut {NoStop}%
\bibitem [{\citenamefont {Tang}\ \emph
  {et~al.}(2018{\natexlab{a}})\citenamefont {Tang}, \citenamefont {Fu},\ and\
  \citenamefont {Ning}}]{Tang:2018er}%
  \BibitemOpen
  \bibfield  {author} {\bibinfo {author} {\bibfnamefont {R.~L.}\ \bibnamefont
  {Tang}}, \bibinfo {author} {\bibfnamefont {X.~X.}\ \bibnamefont {Fu}},\ and\
  \bibinfo {author} {\bibfnamefont {C.~G.}\ \bibnamefont {Ning}},\ }\bibfield
  {title} {\bibinfo {title} {{Accurate electron affinity of Ti and fine
  structures of its anions}},\ }\href@noop {} {\bibfield  {journal} {\bibinfo
  {journal} {The Journal of Chemical Physics}\ }\textbf {\bibinfo {volume}
  {149}},\ \bibinfo {pages} {134304} (\bibinfo {year}
  {2018}{\natexlab{a}})}\BibitemShut {NoStop}%
\bibitem [{\citenamefont {Fu}\ \emph {et~al.}(2017)\citenamefont {Fu},
  \citenamefont {Li}, \citenamefont {Luo}, \citenamefont {Chen},\ and\
  \citenamefont {Ning}}]{Fu:2017ba}%
  \BibitemOpen
  \bibfield  {author} {\bibinfo {author} {\bibfnamefont {X.~X.}\ \bibnamefont
  {Fu}}, \bibinfo {author} {\bibfnamefont {J.~M.}\ \bibnamefont {Li}}, \bibinfo
  {author} {\bibfnamefont {Z.~H.}\ \bibnamefont {Luo}}, \bibinfo {author}
  {\bibfnamefont {X.~L.}\ \bibnamefont {Chen}},\ and\ \bibinfo {author}
  {\bibfnamefont {C.~G.}\ \bibnamefont {Ning}},\ }\bibfield  {title} {\bibinfo
  {title} {{Precision measurement of electron affinity of Zr and fine
  structures of its negative ions}},\ }\href@noop {} {\bibfield  {journal}
  {\bibinfo  {journal} {The Journal of Chemical Physics}\ }\textbf {\bibinfo
  {volume} {147}},\ \bibinfo {pages} {064306} (\bibinfo {year}
  {2017})}\BibitemShut {NoStop}%
\bibitem [{\citenamefont {Chen}\ and\ \citenamefont
  {Ning}(2016)}]{Chen:2016dk}%
  \BibitemOpen
  \bibfield  {author} {\bibinfo {author} {\bibfnamefont {X.~L.}\ \bibnamefont
  {Chen}}\ and\ \bibinfo {author} {\bibfnamefont {C.~G.}\ \bibnamefont
  {Ning}},\ }\bibfield  {title} {\bibinfo {title} {{Accurate electron affinity
  of Pb and isotope shifts of binding energies of Pb {-}}},\ }\href@noop {}
  {\bibfield  {journal} {\bibinfo  {journal} {The Journal of Chemical Physics}\
  }\textbf {\bibinfo {volume} {145}},\ \bibinfo {pages} {084303} (\bibinfo
  {year} {2016})}\BibitemShut {NoStop}%
\bibitem [{\citenamefont {Luo}\ \emph {et~al.}(2016)\citenamefont {Luo},
  \citenamefont {Chen}, \citenamefont {Li},\ and\ \citenamefont
  {Ning}}]{Luo:2016ez}%
  \BibitemOpen
  \bibfield  {author} {\bibinfo {author} {\bibfnamefont {Z.~H.}\ \bibnamefont
  {Luo}}, \bibinfo {author} {\bibfnamefont {X.~L.}\ \bibnamefont {Chen}},
  \bibinfo {author} {\bibfnamefont {J.~M.}\ \bibnamefont {Li}},\ and\ \bibinfo
  {author} {\bibfnamefont {C.~G.}\ \bibnamefont {Ning}},\ }\bibfield  {title}
  {\bibinfo {title} {{Precision measurement of the electron affinity of
  niobium}},\ }\href@noop {} {\bibfield  {journal} {\bibinfo  {journal}
  {Physical Review A}\ }\textbf {\bibinfo {volume} {93}},\ \bibinfo {pages}
  {020501(R)} (\bibinfo {year} {2016})}\BibitemShut {NoStop}%
\bibitem [{\citenamefont {Chen}\ and\ \citenamefont
  {Ning}(2017)}]{Chen:2017ht}%
  \BibitemOpen
  \bibfield  {author} {\bibinfo {author} {\bibfnamefont {X.~L.}\ \bibnamefont
  {Chen}}\ and\ \bibinfo {author} {\bibfnamefont {C.~G.}\ \bibnamefont
  {Ning}},\ }\bibfield  {title} {\bibinfo {title} {{Observation of Rhenium
  Anion and Electron Affinity of Re}},\ }\href@noop {} {\bibfield  {journal}
  {\bibinfo  {journal} {The Journal of Physical Chemistry Letters}\ }\textbf
  {\bibinfo {volume} {8}},\ \bibinfo {pages} {2735} (\bibinfo {year}
  {2017})}\BibitemShut {NoStop}%
\bibitem [{\citenamefont {Tang}\ \emph
  {et~al.}(2018{\natexlab{b}})\citenamefont {Tang}, \citenamefont {Chen},
  \citenamefont {Fu}, \citenamefont {Wang},\ and\ \citenamefont
  {Ning}}]{Tang:2018ct}%
  \BibitemOpen
  \bibfield  {author} {\bibinfo {author} {\bibfnamefont {R.~L.}\ \bibnamefont
  {Tang}}, \bibinfo {author} {\bibfnamefont {X.~L.}\ \bibnamefont {Chen}},
  \bibinfo {author} {\bibfnamefont {X.~X.}\ \bibnamefont {Fu}}, \bibinfo
  {author} {\bibfnamefont {H.}~\bibnamefont {Wang}},\ and\ \bibinfo {author}
  {\bibfnamefont {C.~G.}\ \bibnamefont {Ning}},\ }\bibfield  {title} {\bibinfo
  {title} {{Electron affinity of the hafnium atom}},\ }\href@noop {} {\bibfield
   {journal} {\bibinfo  {journal} {Physical Review A}\ }\textbf {\bibinfo
  {volume} {98}},\ \bibinfo {pages} {020501(R)} (\bibinfo {year}
  {2018}{\natexlab{b}})}\BibitemShut {NoStop}%
\bibitem [{\citenamefont {Lu}\ \emph {et~al.}(2019)\citenamefont {Lu},
  \citenamefont {Tang}, \citenamefont {Fu},\ and\ \citenamefont
  {Ning}}]{Lu:2019df}%
  \BibitemOpen
  \bibfield  {author} {\bibinfo {author} {\bibfnamefont {Y.}~\bibnamefont
  {Lu}}, \bibinfo {author} {\bibfnamefont {R.~L.}\ \bibnamefont {Tang}},
  \bibinfo {author} {\bibfnamefont {X.~X.}\ \bibnamefont {Fu}},\ and\ \bibinfo
  {author} {\bibfnamefont {C.~G.}\ \bibnamefont {Ning}},\ }\bibfield  {title}
  {\bibinfo {title} {{Measurement of the electron affinity of the lanthanum
  atom}},\ }\href@noop {} {\bibfield  {journal} {\bibinfo  {journal} {Physical
  Review A}\ }\textbf {\bibinfo {volume} {99}},\ \bibinfo {pages} {062507}
  (\bibinfo {year} {2019})}\BibitemShut {NoStop}%
\bibitem [{\citenamefont {Wiley}\ and\ \citenamefont
  {McLaren}(1955)}]{Wiley:1955bl}%
  \BibitemOpen
  \bibfield  {author} {\bibinfo {author} {\bibfnamefont {W.~C.}\ \bibnamefont
  {Wiley}}\ and\ \bibinfo {author} {\bibfnamefont {I.~H.}\ \bibnamefont
  {McLaren}},\ }\bibfield  {title} {\bibinfo {title} {{Time-of-Flight Mass
  Spectrometer with Improved Resolution}},\ }\href@noop {} {\bibfield
  {journal} {\bibinfo  {journal} {Review of Scientific Instruments}\ }\textbf
  {\bibinfo {volume} {26}},\ \bibinfo {pages} {1150} (\bibinfo {year}
  {1955})}\BibitemShut {NoStop}%
\bibitem [{\citenamefont {Le{\'o}n}\ \emph {et~al.}(2014)\citenamefont
  {Le{\'o}n}, \citenamefont {Yang}, \citenamefont {Liu},\ and\ \citenamefont
  {Wang}}]{Leon:2014kj}%
  \BibitemOpen
  \bibfield  {author} {\bibinfo {author} {\bibfnamefont {I.}~\bibnamefont
  {Le{\'o}n}}, \bibinfo {author} {\bibfnamefont {Z.}~\bibnamefont {Yang}},
  \bibinfo {author} {\bibfnamefont {H.~T.}\ \bibnamefont {Liu}},\ and\ \bibinfo
  {author} {\bibfnamefont {L.~S.}\ \bibnamefont {Wang}},\ }\bibfield  {title}
  {\bibinfo {title} {{The design and construction of a high-resolution
  velocity-map imaging apparatus for photoelectron spectroscopy studies of
  size-selected clusters}},\ }\href@noop {} {\bibfield  {journal} {\bibinfo
  {journal} {Review of Scientific Instruments}\ }\textbf {\bibinfo {volume}
  {85}},\ \bibinfo {pages} {083106} (\bibinfo {year} {2014})}\BibitemShut
  {NoStop}%
\bibitem [{\citenamefont {Dick}(2014)}]{Dick:2014jf}%
  \BibitemOpen
  \bibfield  {author} {\bibinfo {author} {\bibfnamefont {B.}~\bibnamefont
  {Dick}},\ }\bibfield  {title} {\bibinfo {title} {{Inverting ion images
  without Abel inversion: maximum entropy reconstruction of velocity maps}},\
  }\href@noop {} {\bibfield  {journal} {\bibinfo  {journal} {Phys. Chem. Chem.
  Phys.}\ }\textbf {\bibinfo {volume} {16}},\ \bibinfo {pages} {570} (\bibinfo
  {year} {2014})}\BibitemShut {NoStop}%
\bibitem [{\citenamefont {Fischer}\ \emph {et~al.}(2016)\citenamefont
  {Fischer}, \citenamefont {Godefroid}, \citenamefont {Brage}, \citenamefont
  {J{\"o}nsson},\ and\ \citenamefont {Gaigalas}}]{Fischer:2016gy}%
  \BibitemOpen
  \bibfield  {author} {\bibinfo {author} {\bibfnamefont {C.~F.}\ \bibnamefont
  {Fischer}}, \bibinfo {author} {\bibfnamefont {M.}~\bibnamefont {Godefroid}},
  \bibinfo {author} {\bibfnamefont {T.}~\bibnamefont {Brage}}, \bibinfo
  {author} {\bibfnamefont {P.}~\bibnamefont {J{\"o}nsson}},\ and\ \bibinfo
  {author} {\bibfnamefont {G.}~\bibnamefont {Gaigalas}},\ }\bibfield  {title}
  {\bibinfo {title} {{Advanced multiconfiguration methods for complex atoms: I.
  Energies and wave functions}},\ }\href@noop {} {\bibfield  {journal}
  {\bibinfo  {journal} {Journal of Physics B: Atomic, Molecular and Optical
  Physics}\ }\textbf {\bibinfo {volume} {49}},\ \bibinfo {pages} {182004}
  (\bibinfo {year} {2016})}\BibitemShut {NoStop}%
\bibitem [{\citenamefont {J{\"o}nsson}\ \emph {et~al.}(2013)\citenamefont
  {J{\"o}nsson}, \citenamefont {Gaigalas}, \citenamefont {Biero{\'{n}}},
  \citenamefont {Fischer},\ and\ \citenamefont {Grant}}]{Jonsson:2013ff}%
  \BibitemOpen
  \bibfield  {author} {\bibinfo {author} {\bibfnamefont {P.}~\bibnamefont
  {J{\"o}nsson}}, \bibinfo {author} {\bibfnamefont {G.}~\bibnamefont
  {Gaigalas}}, \bibinfo {author} {\bibfnamefont {J.}~\bibnamefont
  {Biero{\'{n}}}}, \bibinfo {author} {\bibfnamefont {C.~F.}\ \bibnamefont
  {Fischer}},\ and\ \bibinfo {author} {\bibfnamefont {I.~P.}\ \bibnamefont
  {Grant}},\ }\bibfield  {title} {\bibinfo {title} {{New version: Grasp2K
  relativistic atomic structure package}},\ }\href@noop {} {\bibfield
  {journal} {\bibinfo  {journal} {Computer Physics Communications}\ }\textbf
  {\bibinfo {volume} {184}},\ \bibinfo {pages} {2197} (\bibinfo {year}
  {2013})}\BibitemShut {NoStop}%
\bibitem [{\citenamefont {Babushkin}(1964)}]{Babushkin:1964ta}%
  \BibitemOpen
  \bibfield  {author} {\bibinfo {author} {\bibfnamefont {F.~A.}\ \bibnamefont
  {Babushkin}},\ }\href@noop {} {\bibfield  {journal} {\bibinfo  {journal}
  {Acta Physica Polonica}\ }\textbf {\bibinfo {volume} {25}},\ \bibinfo {pages}
  {749} (\bibinfo {year} {1964})}\BibitemShut {NoStop}%
\bibitem [{\citenamefont {J{\"o}nsson}\ \emph {et~al.}(2007)\citenamefont
  {J{\"o}nsson}, \citenamefont {He}, \citenamefont {Froese~Fischer},\ and\
  \citenamefont {Grant}}]{Jonsson:2007dm}%
  \BibitemOpen
  \bibfield  {author} {\bibinfo {author} {\bibfnamefont {P.}~\bibnamefont
  {J{\"o}nsson}}, \bibinfo {author} {\bibfnamefont {X.}~\bibnamefont {He}},
  \bibinfo {author} {\bibfnamefont {C.}~\bibnamefont {Froese~Fischer}},\ and\
  \bibinfo {author} {\bibfnamefont {I.~P.}\ \bibnamefont {Grant}},\ }\bibfield
  {title} {\bibinfo {title} {{The grasp2K relativistic atomic structure
  package}},\ }\href@noop {} {\bibfield  {journal} {\bibinfo  {journal}
  {Computer Physics Communications}\ }\textbf {\bibinfo {volume} {177}},\
  \bibinfo {pages} {597} (\bibinfo {year} {2007})}\BibitemShut {NoStop}%
\bibitem [{\citenamefont {Grant}(1974)}]{Grant:1974cr}%
  \BibitemOpen
  \bibfield  {author} {\bibinfo {author} {\bibfnamefont {I.~P.}\ \bibnamefont
  {Grant}},\ }\bibfield  {title} {\bibinfo {title} {{Gauge invariance and
  relativistic radiative transitions}},\ }\href@noop {} {\bibfield  {journal}
  {\bibinfo  {journal} {Journal of Physics B: Atomic and Molecular Physics}\
  }\textbf {\bibinfo {volume} {7}},\ \bibinfo {pages} {1458} (\bibinfo {year}
  {1974})}\BibitemShut {NoStop}%
\bibitem [{Note1()}]{Note1}%
  \BibitemOpen
  \bibinfo {note} {See Supplemental Material at http://link.aps.org for details
  for the calculations of all transitions and detailed assignments of observed
  peaks.}\BibitemShut {Stop}%
\bibitem [{\citenamefont {Engelking}\ and\ \citenamefont
  {Lineberger}(1979)}]{Engelking:1979gf}%
  \BibitemOpen
  \bibfield  {author} {\bibinfo {author} {\bibfnamefont {P.~C.}\ \bibnamefont
  {Engelking}}\ and\ \bibinfo {author} {\bibfnamefont {W.~C.}\ \bibnamefont
  {Lineberger}},\ }\bibfield  {title} {\bibinfo {title} {{Laser photoelectron
  spectrometry of Fe{-}: The electron affinity of iron and the "nonstatistical"
  fine-structure detachment intensities at 488 nm}},\ }\href@noop {} {\bibfield
   {journal} {\bibinfo  {journal} {Physical Review A}\ }\textbf {\bibinfo
  {volume} {19}},\ \bibinfo {pages} {149} (\bibinfo {year} {1979})}\BibitemShut
  {NoStop}%
\end{thebibliography}%

\end{document}